# Numerical thermalization in 2D PIC simulations: Practical estimates for low temperature plasma simulations

Sierra Jubin[*,1,2] Andrew Tasman Powis,[2] WillcaVillafana,[2] Dmytro Sydorenko,[3] Shahid Rauf,[4] Alexander V. Khrabrov,[2] Salman Sarwar,[1] Igor D. Kaganovich[2]

[1] Princeton University, Princeton, New Jersey 08544, USA

[2] Princeton Plasma Physics Laboratory, Princeton, New Jersey 08543, USA

[3] University of Alberta, Edmonton, Alberta T6G 2E1, Canada

[4] Applied Materials, Inc., 3333 Scott Blvd, Santa Clara, California 95054, USA

*Corresponding Author: sjubin@princeton.edu

The process of numerical thermalization in particle-in-cell (PIC) simulations has been studied extensively. It is analogous to Coulomb collisions in real plasmas, causing particle velocity distributions (VDFs) to evolve towards a Maxwellian as macroparticles experience polarization drag and resonantly interact with the fluctuation spectrum. This paper presents a practical tutorial on the effects of numerical thermalization in 2D PIC applications. Scenarios of interest include simulations which must be run for many thousands of plasma periods and contain a population of cold electrons that leave the simulation space very slowly. This is particularly relevant to many low temperature plasma discharges and materials processing applications. We present numerical drag and diffusion coefficients and their associated timescales for a variety of grid resolutions, discussing the circumstances under which the electron VDF is modified by numerical thermalization. Though the effects described here have been known for many decades, direct comparison of analytically derived, velocity-dependent numerical relaxation timescales to those of other relevant processes has not often been applied in practice due to complications that arise in calculating thermalization rates in 1D simulations. Using these comparisons, we estimate the impact of numerical thermalization in several example low temperature plasma applications including capacitively coupled plasma (CCP) discharges, inductively coupled plasma (ICP) discharges, beam plasmas, and hollow cathode discharges. Finally, we discuss possible strategies for mitigating numerical relaxation effects in 2D PIC simulations.

## 1. Introduction

The particle-in-cell (PIC) method of plasma simulation is frequently used for simulations of plasma discharges in plasma processing. [1, 2, 3, 4, 5] The importance of avoiding artificial numerical effects caused by the increased density fluctuations inherent in the PIC method is often insufficiently analyzed. The main focus of this paper is to provide accurate theoretical estimates of the effect of numerical noise on the Electron Velocity Distribution Function (EVDF) and give examples of such analysis for widely used discharges in plasma processing: capacitively and inductively-coupled plasmas, hollow cathode plasmas, and electron beam-generated plasmas.

In most PIC simulations of low temperature plasmas the constraint on the number of macroparticles required to achieve accuracy using the particle-in-cell (PIC) method of plasma simulation is often expressed more qualitatively than quantitatively, prescribing that the number of macroparticles per cell should be 'large' (several hundred or several thousand). Convergence tests are recommended to ensure that the results of a simulation do not depend on the number of macroparticles used. While such convergence tests are important, they should be aided by a more rigorous analysis of the nature of the error caused by numerical noise. The theory describing the effect of numerical noise on the EVDF has been developed decades ago. However, it is not often used due to the so-called "kinetic blocking" effects that reduce the effect of numerical noise in 1D PIC simulations. [6, 7] The rate of numerical thermalization is often significantly slower in 1D than in 2D and analytical prediction of the relaxation time is difficult in 1D. By contrast, the theory is relatively simple in multidimensional PIC simulations and the thermalization rates are often much faster. [8, 9] Until recently most PIC simulations were 1D and hence the effects of noise were not of great concern. The situation is drastically different in 2D where, as we show below, the effects of noise and the corresponding thermalization can strongly affect the EVDF. Hence, these effects need to be carefully evaluated.

The results presented here are a pedagogical synthesis of the available information for practical application to 2D PIC simulations. We are particularly focused on PIC simulations of low temperature plasma discharges, where fluid models are frequently used under the assumption that particle velocity distribution functions (VDFs) are already Maxwellian. The use of a kinetic simulation technique such as the PIC method implies that there are possible kinetic effects to be investigated. However, without adequate attention to numerical thermalization errors, the potential features of interest in the electron velocity distribution function (EVDF) – such as an excess or depleted population of high velocity electrons – will not be accurately represented.

Numerical thermalization drives particle VDFs towards a Maxwellian, with EVDFs thermalizing more rapidly than ion velocity distribution functions (IVDFs). This thermal relaxation often happens on a timescale ($\tau_R$) much smaller than the numerical heating timescale ($\tau_H$) as can be seen in the example 2d3v (two spatial dimensions; three velocity dimensions) PIC simulation of a spatially uniform plasma of density $1.24 \times 10^{14}\ m^{-3}$ in Fig. 1. Here we used the explicit leapfrog time integration method with an explicit momentum conserving scheme to simulate the plasma in a 128 by 128 cell cartesian grid, with periodic boundary conditions, a grid spacing of 0.23 mm, and 40 particles per cell. We observe that an initial waterbag EVDF

$$f(v_x, v_y, v_z) = \Pi(v_x)\Pi(v_y)\Pi(v_z), \qquad \Pi(v) = \begin{cases} \frac{1}{2v_c}, & -v_c < v < v_c \\ 0, & v > v_c\ |\ v < -v_c \end{cases} \qquad (1)$$

with a cutoff velocity $v_c$ corresponding to the velocity of an electron with 4.5 eV of kinetic energy experiences numerical thermalization within a few hundred plasma frequency periods. By comparison, the average electron kinetic energy only increases by ~2.5% after ten thousand plasma frequency periods. The numerical thermalization results in a rapid increase in the kinetic entropy calculated from the one-dimensional EVDFs in the x- and y-direction. (See Fig. 1B.) Although the difference between the kinetic entropy of a Maxwellian distribution and the kinetic entropy of a waterbag distribution is small relative to their absolute value, it provides a very clear signal as it is obtained from the global EVDF of the entire simulation domain. Further information on thermalization measurement techniques can be found in Section 3.1.

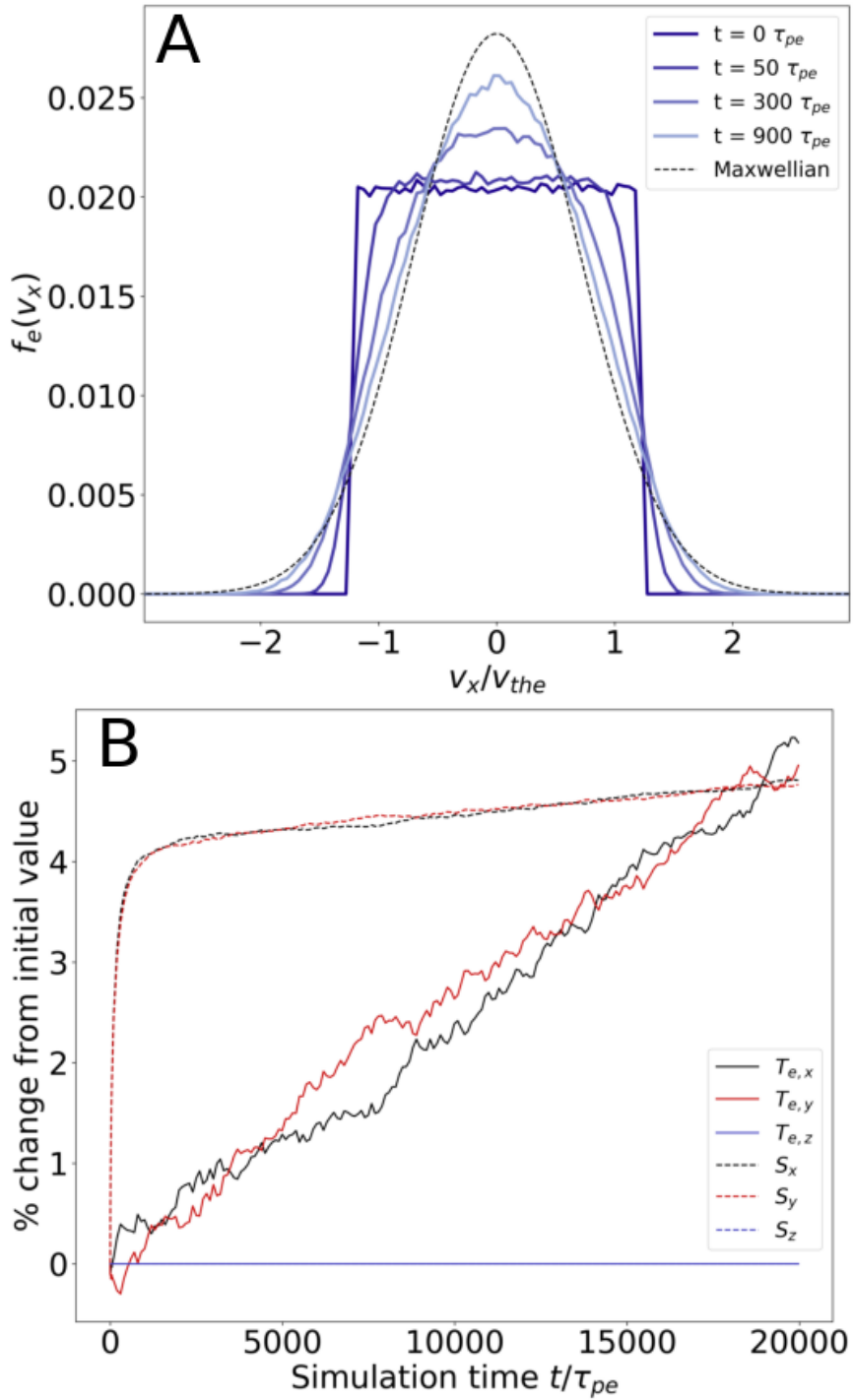

**Figure 1**: Numerical thermalization of an EVDF for a 2d3v electrostatic PIC simulation run using the PIC code EDIPIC-2D *[10]*. A) The evolution of the one-dimensional EVDF in the x-direction as it approaches a Maxwellian, and B) the "one-dimensional" electron temperature $T_{e,j}$ and kinetic entropy $S_j$ in the $j^{th}$ dimension calculated from the one-dimensional EVDFs as shown in Eqns 2 and 3. The dashed blue line and solid blue line representing the quantities in the spatially unrepresented z-dimension lie on top of each other.

The quantities shown in Fig. 1B, denoted $T_{e,j}$ and $S_j$, respectively, are one-dimensional measurements of electron temperature and a quantity related to the entropy of the normalized electron distribution function, $f(\mathbf{v})$.

$$T_{e,j} = m_e \int dv_j \, v_j^2 f(v_j) \tag{2}$$

$$S_j = -k_B \int dv_j \, f(v_j) \ln(f(v_j)) \tag{3}$$

They are defined while considering the EVDF only as a function of one velocity dimension $j \in \{x, y, z\}$. This means that the various $S_j$ cannot be summed to produce an actual measurement of the kinetic entropy of the full multidimensional EVDF. However, separately tracking this measurement in each dimension illustrates that without Monte Carlo collisions or magnetic fields which might redirect particle motion out of the simulated spatial plane, the evolution of the EVDF in a 2d3v simulation only occurs in velocity dimensions for which the electric field is defined. Thus, for the 2d3v simulations depicted in Fig. 1, the one-dimensional distribution function in the z-direction does not evolve even after twenty thousand plasma frequency periods have passed. This of course will not be the case for 3D PIC, where the ability of particles to extract energy from fields in all three spatial dimensions makes it an attractive tool for modeling kinetic turbulence. [11]

Numerical thermalization occurs due to a process which is analogous to Coulomb collisions in real plasmas and has alternately been described as "numerical collisions." The basic form of a numerical collision operator is plainly stated in Birdsall and Langdon's textbook [12]. It has also recently been rederived including the effects of multiple particle species with lucid and detailed discussion by Touati et al. [13].

Because numerical collisions are caused by the inherent graininess of particle-based simulations, they do not disappear in the limit of infinitely small grid spacing and timestep. The numerical collision operator is a correction to the Vlasov evolution of plasma in phase space, and to first order it scales with the inverse of the numerical plasma parameter $1/N_{Dmac}$, where $N_{Dmac}$ is the number of macroparticles per Debye volume. For our purposes this is defined as $N_{Dmac} = n_{mac}\lambda_D$ in a 1D simulation, $N_{Dmac} = n_{mac}\lambda_D^2$ in a 2D simulation, and $N_{Dmac} = n_{mac}\lambda_D^3$ in a 3D simulation with $\lambda_D$ representing the Debye length and $n_{mac}$ the macroparticle number density. Cancellation of a large portion of this first order term in 1D PIC plasmas can cause the rate of numerical thermalization to depend on a second order term which scales as $1/N_{Dmac}^2$, a process known as kinetic blocking [7]. The kinetic blocking can be disrupted by the addition of Monte Carlo collisions to the simulation, recovering a first order term [14, 15, 16].

The numerical collision operator acting on species α can be written in a Fokker-Planck form representing drag $A_\alpha(\mathbf{v})$ and diffusion $D_\alpha(\mathbf{v})$ in velocity space.

$$\left(\frac{\partial f_\alpha}{\partial t}\right)_c = -\frac{\partial}{\partial v}\cdot[A_\alpha(\mathbf{v})f_\alpha(\mathbf{v})] + \frac{1}{2}\frac{\partial}{\partial v}\frac{\partial}{\partial v}:[D_\alpha(\mathbf{v})f_\alpha(\mathbf{v})] \qquad (4)$$

$A_\alpha(\mathbf{v})$ is a vector parallel to the direction of v , while $D_\alpha(\mathbf{v})$ can be written as a diagonal matrix in a velocity space defined by a basis set of vectors parallel and perpendicular to v. Drag is caused both by the polarization which occurs when Debye shielding around a moving macroparticle lags behind it and also by the transfer of kinetic energy from the moving macroparticle into the energy of the fluctuating electric field. Diffusion is similarly caused by resonant interactions with these fluctuations; the diffusion coefficient is proportional to an integral over the fluctuation spectrum. This is identical to a description of Coulomb collisions between charged particles in real plasmas, and this is because numerical thermalization can be thought of as Coulomb collisions between macroparticles. Velocity-dependent relaxation rates arise directly from the drag and diffusion coefficients of the Fokker-Planck form.

There are, however, three main differences that distinguish numerical thermalization from the effects of Coulomb collisions between real particles. Firstly, the fluctuation spectrum is modified by aliasing effects due to finite grid resolutions and timesteps. Second, interactions between particles are reduced by the use of shape functions which give macroparticles finite width. The final possible difference is the reduced number of dimensions, which has a significant impact on 1D simulations as described above.

This paper is divided into five sections. In Section 2 we will present the drag and diffusion coefficients for a test particle electron in a Maxwellian electron background for 2D electrostatic explicit PIC simulations using a cloud-in-cell (CIC) particle weighting scheme on a uniform cartesian grid. These coefficients will be compared to the drag and diffusion coefficients of the Landau collision operator. In Section 3 we will then discuss the relaxation rates related to these coefficients and examine their resemblance to a measured empirical scaling of numerical thermalization time. We will also discuss possible strategies for reducing the rate of numerical thermalization. In Section 4 we will assess the relaxation rates in application to various 2D PIC simulations of low temperature plasmas, including capacitively coupled plasma (CCP) discharges, inductively coupled plasma (ICP) discharges, electron beam generated plasmas, and hollow cathode discharges. Finally, we will summarize the results and give final recommendations for addressing numerical thermalization in 2D PIC simulations.

## 2. Numerical thermalization drag and diffusion coefficients in 2D PIC

We first present the drag and diffusion coefficients from the Fokker-Planck form of the 2D electrostatic PIC numerical collision operator for test particle electrons in a homogeneous quasineutral background plasma with one singly charged ion species. This analysis is carried out for simulations using a uniform, effectively infinite cartesian grid. Both background electrons and ions are assumed to be Maxwellian, with respective temperatures $T_e$ and $T_i$ and masses $m_e$ and $m_i$. The numerical collision operator was calculated from the results derived by Touati et al [13] for explicit PIC simulations, where we have neglected the effects of finite time differencing. This is done under the assumption that the time step is sufficiently well-resolved with respect to the plasma frequency so that time aliasing modes other than the zeroth mode are negligible. We have also assumed equal macroparticle weighting factors for electrons and ions, resulting in an equal density of electron and ion macroparticles. Further details regarding the calculation of these Fokker-Planck (Eqn. 4) drag and diffusion coefficients can be found in Appendix A.

$$A^e = \sum_\beta A^{e/\beta}, \qquad \boldsymbol{D}^e = \sum_\beta \boldsymbol{D}^{e/\beta}, \qquad D_\parallel^e = \widehat{\boldsymbol{V}}^T \cdot \boldsymbol{D}^e \cdot \widehat{\boldsymbol{V}}, \qquad D_\perp^e = \widehat{\boldsymbol{V}}_\perp^T \cdot \boldsymbol{D}^e \cdot \widehat{\boldsymbol{V}}_\perp \qquad (5)$$

$$A^{e/\beta} = -\frac{\omega_{pe} v_{Te}}{N_{Dmac}} I_A^{e/\beta}, \qquad \boldsymbol{D}^{e/\beta} = \frac{\omega_{pe} v_{Te}^2}{N_{Dmac}} \boldsymbol{I}_D^{e/\beta} \qquad (6)$$

$$I_A^{e/\beta} = \frac{1+m_e/m_\beta}{2^{5/2}\pi^{3/2}R_{v\beta}^3} \int d\boldsymbol{k} \sum_p \left|\frac{S(\boldsymbol{k})S(\boldsymbol{k}_p)}{\epsilon(\boldsymbol{k}\cdot\boldsymbol{V},\boldsymbol{k})\mathcal{K}^2}\right|^2 \frac{(\mathcal{K}\cdot\boldsymbol{k}_p)(\boldsymbol{k}\cdot\boldsymbol{V})(\mathcal{K}\cdot\widehat{\boldsymbol{V}})}{k_p^3} e^{-(\boldsymbol{k}\cdot\boldsymbol{V})^2/k_p^2 R_{v\beta}^2} \qquad (7)$$

$$I_D^{e/\beta} = \frac{1}{2^{5/2}\pi^{3/2}R_{v\beta}} \int d\pmb{k} \sum_p \left|\frac{S(\pmb{k})S(\pmb{k}_p)}{\epsilon(\pmb{k}\cdot V,\pmb{k})\mathcal{K}^2}\right|^2 \frac{(\mathcal{K}\otimes\mathcal{K})}{k_p} e^{-(\pmb{k}\cdot V)^2/k_p^2 R_{v\beta}^2} \tag{8}$$

$$\epsilon(\pmb{k}\cdot V, \pmb{k}) \equiv 1 - \sum_p \frac{S(\pmb{k}_p)^2(\mathcal{K}\cdot\pmb{k}_p)}{2\mathcal{K}^2 k_p^2}\left[Z'\left(\frac{\pmb{k}\cdot V}{k_p}\right) + \frac{T_e}{T_i}Z'\left(\frac{\pmb{k}\cdot V}{k_p R_{v\beta}}\right)\right] \tag{9}$$

These coefficients are proportional to dimensionless integrals presented as functions of scaled velocity $V = \mathbf{v}/\mathrm{v}_{Te}$ where $\mathrm{v}_{Te} = \sqrt{2k_B T_e/m_e}$ is the thermal velocity of the background electrons. The test particle interactions with species $\beta = (i, e)$ are also given as a function of the ratio of the species' thermal velocity with that of the background electrons, so $R_{v\beta} = \mathrm{v}_{T\beta}/\mathrm{v}_{Te} = \sqrt{T_\beta m_e/T_e m_\beta}$. Wavenumbers and similar quantities have been scaled by the electron Debye length (i.e. $\pmb{k} = k\lambda_{De}$ with $\lambda_{De} = \sqrt{k_B T_e \varepsilon_0/nq^2}$) so that the integral over the volume $d\pmb{k} = d\pmb{k}_x d\pmb{k}_y$ is a dimensionless quantity. The spatial aliasing effects are accounted for through sums over all integers $\pmb{p} = (p_x, p_y)$, which couple aliased wavelengths with corresponding wavenumbers dependent on the spatial grid resolution, $\pmb{k}_p = \pmb{k} - \pmb{p}\, 2\pi/R_g$ where $R_g = \Delta_x/\lambda_{De}$ is the grid spacing $\Delta_x$ scaled by the electron Debye length. Here we assume $\Delta_x$ to be equal in both directions. $\mathcal{K}$ is a quantity dependent on the particular Maxwell solver used, but here assumed to be equal to the scaled wavenumber $\pmb{k}$ as would be the case for a spectral Maxwell solver. For our calculations we have chosen the commonly used CIC method for assigning charge to the grid, yielding a shape function with the Fourier transform $S(\pmb{k}) = \mathrm{sinc}(\pmb{k}_x R_g/2)^2 \mathrm{sinc}(\pmb{k}_y R_g/2)^2$. Here $\mathrm{sinc}(x) = \frac{\sin(x)}{x}$.

As in a real plasma, the contribution to the dielectric function $\epsilon(\pmb{k}\cdot V, \pmb{k})$ from each species $\gamma$ is proportional to its plasma frequency squared: $\omega_{p\gamma}^2 = n_\gamma q_\gamma^2/\varepsilon_0 m_\gamma$. Also in the dielectric function is the plasma dispersion function $Z(x)$, where $Z'(x)$ is its derivative evaluated at $x$. We choose not to approximate $\epsilon(\boldsymbol{k} \cdot \boldsymbol{V}, \boldsymbol{k})$ with its static approximation at $\boldsymbol{V} = 0$ based on the results of Okuda and Birdsall [17]. They showed that the static shielding approximation (Landau collision operator analogue) is not a good approximation of the full Balescu-Lenard collision operator analogue for collisions between clouds of charge or for particles in a reduced number of dimensions.

A key feature to note is that both the drag and diffusion coefficients are inversely proportional to $N_{Dmac} = n_{mac}\lambda_D^2 = N_{ppc}/R_g^2$ where $N_{ppc}$ is the number of particles per cell. Unlike real Coulomb collisions in 3D, there is no logarithmic dependence on $N_{Dmac}$ in the numerator. This is because, thanks to the shape functions, there is no need to truncate the integrals at large $\boldsymbol{k}$ values (small wavelengths) in order to achieve convergence.

The numerical drag and diffusion coefficients for a Maxwellian electron background can be compared (see Fig. 2) to drag and diffusion coefficients for the Landau collision operator for a test electron in a Maxwellian electron background:

$$A_L^{e/e} = \frac{\omega_{pe} v_{Te} \ln \Lambda_{ee}}{2 N_D} I_{A,L}^{e/e}, \qquad \boldsymbol{D}_L^{e/e} = \frac{\omega_{pe} v_{Te}^2 \ln \Lambda_{ee}}{2 N_D} \boldsymbol{I}_{D,L}^{e/e} \tag{10}$$

$$I_{A,L}^{e/e} = \frac{1}{2^{1/2}\pi} \left[ \frac{erf(V_e) - 2\pi^{-1/2} V_e e^{-V_e^2}}{2 V_e^2} \right] \tag{11}$$

$$I_{D\|,L}^{e/e} = \frac{1}{2^{3/2}\pi} \left[ \frac{erf(V_e) - 2\pi^{-1/2} V_e e^{-V_e^2}}{2 V_e^3} \right] \tag{12}$$

$$I_{D\perp,L}^{e/e} = \frac{1}{2^{5/2}\pi} \left[ \frac{2V_e^2 \, erf(V_e) - erf(V_e) + 2\pi^{-1/2} V_e e^{-V_e^2}}{2 V_e^3} \right] \tag{13}$$

Here $V_e = v/v_{Te}$ is the velocity scaled by the electron thermal velocity, $N_D = n\lambda_D^3$ is the number of real particles per Debye cube, and $\ln \Lambda_{ee} = \ln(4\pi N_D)$ is the Coulomb logarithm. The numerical drag and diffusion coefficients depend not only on the magnitude of $\boldsymbol{V_e}$ but also its orientation with respect to the grid. However, this dependence was found to be negligible when grid spacings were set equal in both directions, and in Fig. 2 the orientation of the test particle velocity was taken to be along one of the grid axes.

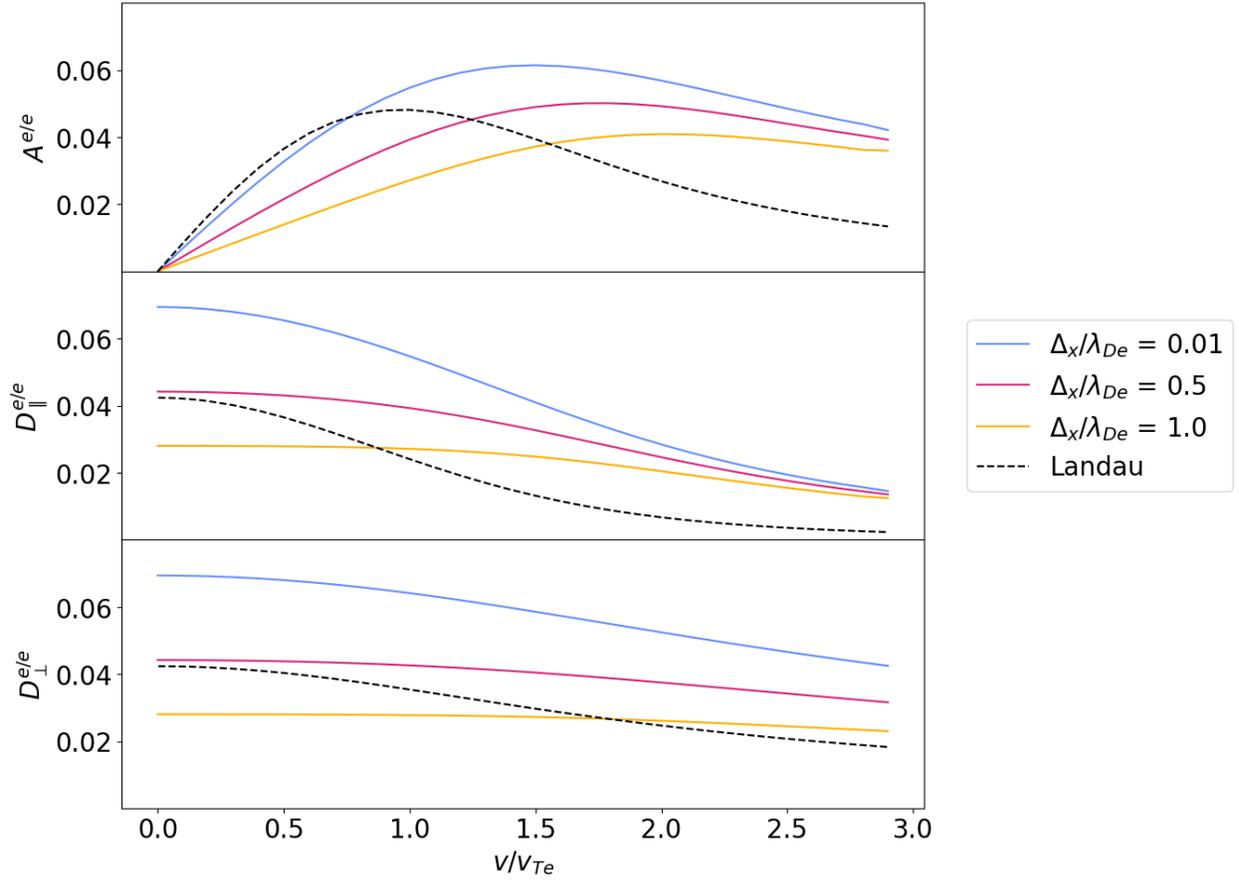

**Figure 2**: Scaled numerical drag and diffusion coefficients at a variety of grid mesh sizes (solid color lines) and drag and diffusion coefficients for the Fokker-Planck form of the Landau collision operator (dashed black lines) for test particle electron interactions with Maxwellian background electrons. Numerical drag coefficients are scaled by $\omega_{pe} v_{Te}/N_{Dmac}$ and the corresponding diffusion coefficients are scaled by $\omega_{pe} v_{Te}^2/N_{Dmac}$ while the drag

coefficients of the Landau collision operator are scaled by $\omega_{pe} v_{Te} \ln \Lambda_{ee} / 2N_D$ and the corresponding diffusion coefficients are scaled by $\omega_{pe} v_{Te}^2 \ln \Lambda_{ee} / 2N_D$.

The coefficients shown in Fig. 2 can be compared to those derived by Abraham-Shrauner [18] for single species point particles in 2D with a Lorentz background distribution, those derived by Okuda and Birdsall [17] for plasmas of single species Gaussian charge clouds in 2D, and those derived by Reynolds et al [19] for multispecies point particles in 2D. Similar to real Coulomb collisions, the drag coefficient is linear at $v \ll v_{Te}$. The parallel and perpendicular diffusion coefficients are equal at $v = 0$, with perpendicular diffusion (corresponding to pitch angle scattering) falling off more slowly as the test particle velocity increases.

As shown by Okuda and Birdsall, increasing the width of the particle shape functions (here accomplished by increasing the size of the grid spacing) reduces numerical drag and diffusion. Decreasing the width of the particle shape function causes the numerical drag and diffusion coefficients to approach that of a point particle [17]. However, at $v \gg v_{Te}$, the effect of the finite macroparticle width is reduced. Okuda and Birdsall attribute this to the fact that fast moving charge clouds resonantly interact with background fluctuations at wavelengths which scale linearly with the test particle velocity; thus, there is some $v$ at which the wavelength of the resonant interaction is larger than the width of the shape function and the behavior of the charge cloud is indistinguishable from that of a point particle. [17]

We can see in Fig. 2 that drag and diffusion coefficients of the Landau collision operator appear to fall off much more rapidly with increasing velocity than the 2D numerical drag and diffusion coefficients. However, the behavior of the 2D coefficients at very large velocities has been shown to have the same dependence on v as the coefficients in 3D in the limit that $v \gg v_T$. [19]

## 3. Numerical thermalization rates and mitigation strategies

### 3.1 Relaxation processes and rates

From these coefficients we can define the three relaxation processes and their corresponding rates as they are usually written: slowing down, parallel diffusion, and perpendicular diffusion.

$$\nu_s^{e/\beta} = \frac{A^{e/\beta}}{v}, \quad \nu_\parallel^{e/\beta} = \frac{D_\parallel^{e/\beta}}{v^2}, \quad \nu_\perp^{e/\beta} = \frac{D_\perp^{e/\beta}}{v^2} \qquad (14)$$

It is immediately apparent that any thermalization time based on these processes will scale with $\tau_R \sim N_{D,mac}/\omega_{\text{pe}}$. While there is no relaxation timescale which universally applies to every initial electron distribution function, there are nevertheless many methods which have been used to measure some definition of $\tau_R$ in PIC plasmas.

Many of these measurement methods involve setting up an initial waterbag EVDF and tracking its evolution via a parameter which measures the distribution's resemblance to a Maxwellian. Examples of these parameters include the value of the velocity distribution function at zero velocity [20], a measure of the kinetic entropy of the system [9, 21, 22], the coefficients of an expansion of the distribution function in eigenfunctions of the Lenard-Bernstein collision operator [23, 14, 15], and a simple standard statistical chi-squared fit to a Maxwellian [15, 16].

The thermalization time has also often been equated to the slowing-down timescale ($\tau_s$) associated with the drag on a slow-moving test particle. [24, 8] This resulted in initial errors in estimating the thermalization time in a 1D single-species plasma [24] due to the necessity of distinguishing between drag and diffusion timescales for a test particle ($\tau_s, \tau_\parallel, \tau_\perp \sim N_{Dmac}$) and processes which evolve the total velocity distribution function ($\tau_R \sim N_{Dmac}^2$). There is no such distinction in 2D, and so the scaling of this test particle slowing-down time will accurately reflect the scaling of the thermalization time ($\tau_s, \tau_R \sim N_{Dmac}$) [8].

As mentioned before, complications arise in 2D PIC simulations if there is an applied magnetic field which points out of the plane of simulation. Numerical collisionality is enhanced as the gyromotion of particles can cause Coulomb interactions between particles to repeat in a reinforcing manner. The inability of the particles to travel away from each other along a nonexistent third dimension parallel to the magnetic field lines can cause the scaling of the thermalization time under these circumstances to be $\tau_R \sim \sqrt{N_{Dmac}}$ [25, 21, 22].

We can compare our analytical estimates for the average test particle drag timescale in a 2D PIC plasma with a CIC shape function to Hockney's [8] empirical estimate of the same:

$$\bar{\tau}_s^{Hockney} = \tau_{pe} \frac{N_{Dmac}}{K_1} \left[1 + \left(\frac{\Delta_x}{\lambda_D}\right)^2\right], \quad K_1 = 0.98 \pm 0.20 \qquad (15)$$

Considering only drag due to electron-electron interactions, we can calculate this average drag timescale by integrating over velocity:

$$\bar{\tau}_s^{e/e} = \int \frac{d\mathbf{v}}{v_s^{e/e}} \frac{1}{\pi v_{Te}^2} e^{-v^2/v_{Te}^2} \qquad (16)$$

The comparison in Fig. 3 shows that this is reasonably close to Hockney's empirical estimate, which provides an order of magnitude estimate for the timescale of numerical thermalization.

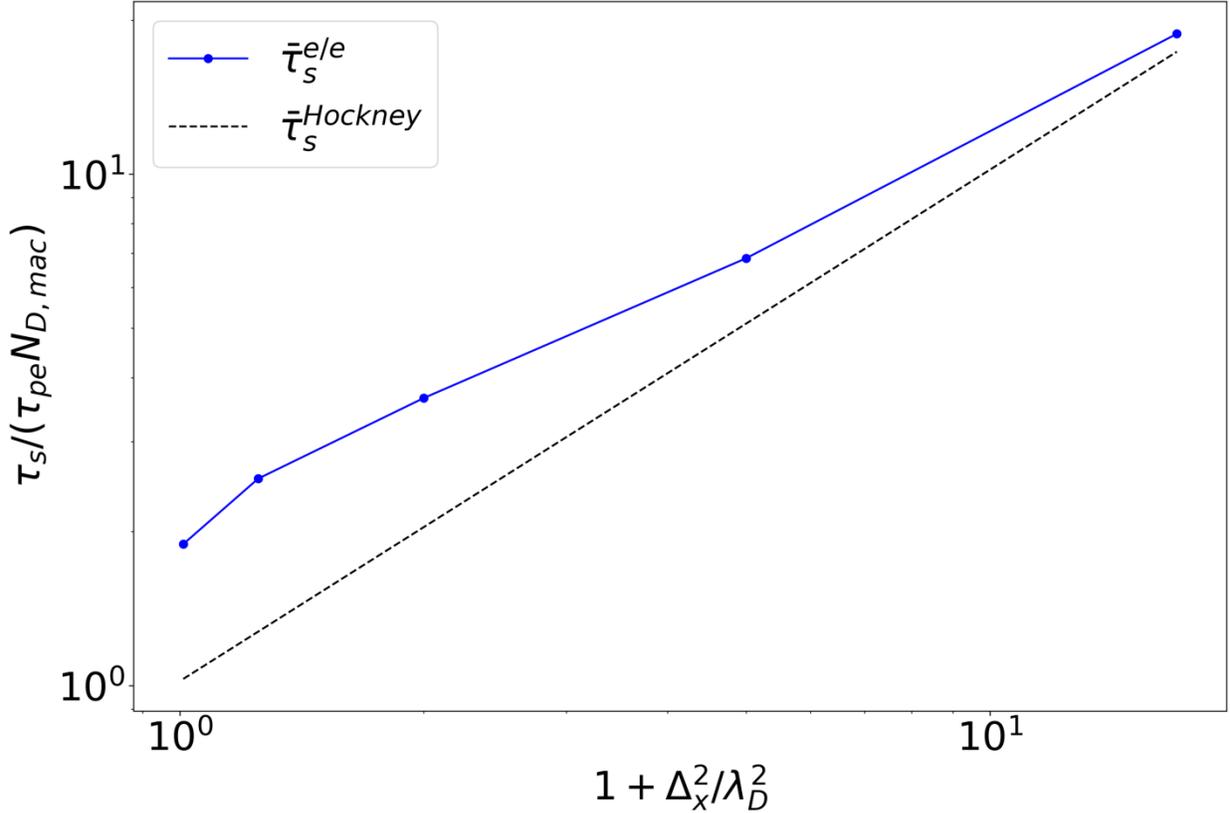

**Figure 3**: Numerical collision timescales for 2D PIC with a CIC scheme. Comparison of timescales derived from analytical theory (blue solid curve) with the empirical scaling given by Hockney (black dashed curve) *[8]*. Hockney's estimates accounted for large grid spacings up to 32 times the Debye length.

We should note that Hockney's method [8] of analyzing numerical collisional time scales by conducting "numerical experiments" to measure average electron slowing-down and deflection times is an invaluable test for any new particle-in-cell code. In cases where a numerical collision operator cannot be easily analytically defined direct measurement is especially useful. Simulations with unstructured grids fall into this category, and measurement of slowing down and deflection times in these cases offer the best insight into numerical thermalization rates. Examples of this type of analysis for unstructured grids can be found in work by Gatsonis and Spirkin [26] and work by Averkin and Gatsonis [27].

### 3.2 Mitigation strategies

Given that numerical thermalization happens on timescales roughly comparable to $N_{Dmac}\tau_{pe}$, the most obvious way to delay the onset of numerical thermalization is to increase the number density of macroparticles. Increasing the grid spacing size with respect to the Debye length produces effectively larger macroparticle shape functions, reducing the Coulomb interactions between macroparticles and delaying numerical thermalization. Significant reductions in thermalization time are only achieved by under-resolving the Debye length, however, and may result in numerical instabilities in explicit schemes.

Implicit or energy conserving schemes [28] can be used to create larger grid sizes while avoiding numerical heating due to grid aliasing, but one should note that the thermalization rate depends linearly on the *density* of the macroparticles, so that simulations with large grid length scales should use larger numbers of macroparticles per cell to fully take advantage of the reduced numerical collisions offered by broad shape functions. The numerical collision operators shown in Section 2 strictly speaking only apply to the explicit schemes for which they were derived, but as can be seen in Fig. 4A, the numerical thermalization rates are comparable for a direct implicit and explicit scheme with a well-resolved grid resolution, timestep, and number of particles per cell. There we can also see that increasing the grid length scale above the Debye length while maintaining the same density of macroparticles will result in a reduced rate of numerical thermalization. Numerical heating (or cooling) in the direct implicit scheme with an under-resolved Debye length is not ignorable, but it can be made negligible using a judicious choice of timestep; according to Sun et al [29] this "optimal path" of near-zero numerical heating for direct implicit algorithms occurs when the ratio between $\omega_{pe} dt$ and $\Delta_x/\lambda_{De}$ is roughly 0.1. As can be seen in Fig. 4B, the direct implicit simulation with a grid spacing of $\Delta_x = 3\lambda_{De}$ does experience a modest ~6% increase in the kinetic energy of the electrons over ten thousand plasma periods – if stricter control of heating is required the time step can be refined further.

The simulated densities and initial EVDFs are identical to those described in Fig. 1. The explicit and direct implicit simulations with a well-resolved Debye length ($\Delta_x = 0.2\lambda_D$) contain 40 particles per cell while the direct implicit case with an unresolved Debye length ($\Delta_x = 3\lambda_D$) contains 9000 particles per cell, leaving the density of macroparticles constant between the cases. The setup and the quantities calculated are identical to those discussed in Section 1, though the one-dimensional kinetic entropy in Fig 4A is presented as a fractional value of the entropy of a Maxwellian with the average kinetic energy of the EVDF at that point in time. In this way, we compare increases in entropy due to numerical thermalization more directly, ignoring increases in entropy due to numerical heating.

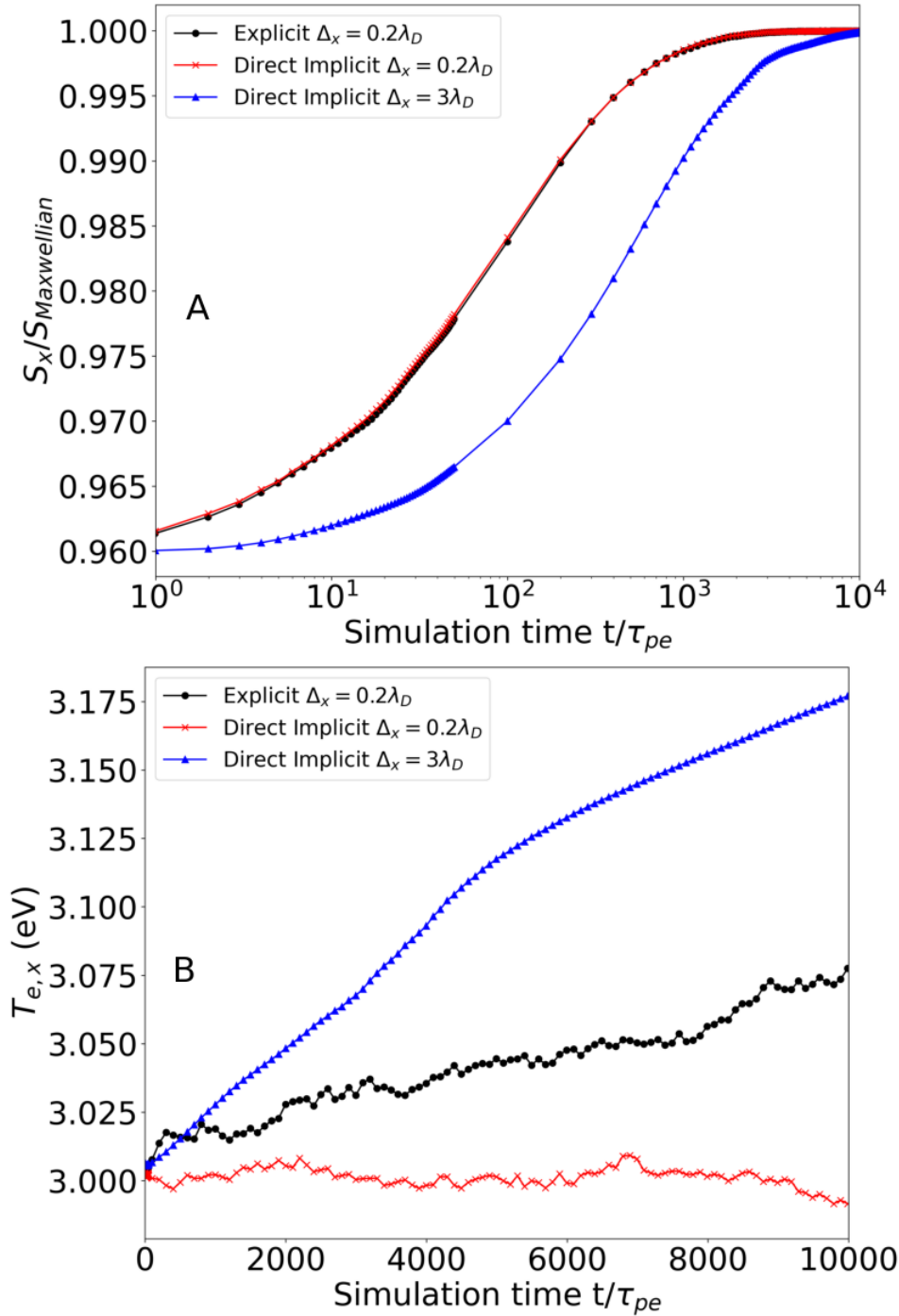

**Figure 4:** The evolution of the "one-dimensional" electron kinetic entropy (A) and electron temperature (B) in the x-direction (see Eqns. 2 and 3) for explicit and direct implicit PIC simulations in EDIPIC-2D with initial waterbag EVDFs.

Aside from using implicit schemes with large grid spacings, another option which effectively increases the width of the macroparticle charge clouds and thus delays thermalization is to use an interpolation scheme of a higher order than the CIC scheme analyzed in this work. There is also some indication that unstructured grids may result in reduced thermalization times; the slowing-down and deflection times measured by Averkin and Gatsonis for the 3D electrostatic PIC code EUPIC with unstructured teterahedral grids were in fact larger than the measured heating times [27]. This should prompt further investigation into the use of such grids as a potential avenue to minimize numerical thermalization.

Finally, for some applications it is reasonable to reduce the effective plasma frequency by scaling the permittivity of free space, $\varepsilon_0$, by some large factor. The latter strategy will, of course, only be useful in situations where there is certainty that other important phenomena such as instabilities are of little interest. Many low temperature plasma sources fall into this category, and in some of these cases it is possible to sufficiently reduce the numerical collision rate below that of real Coulomb collisions by increasing $\varepsilon_0$. See for example the hollow cathode simulations detailed in Section 4.2.

A simple glance at the scaling in Fig. 2 is sufficient to infer that reducing the effects of numerical thermalization below that of real Coulomb collisions by simply increasing the number of macroparticles is often computationally expensive to attempt. To make the numerical processes in 2D PIC roughly comparable to the drag and diffusion caused by true Coulomb collisions (at velocities small compared to the electron thermal velocity), it would require that a macroparticle represent only a handful of real particles: $N_{Dmac} \approx 2N_D/ln\, \Lambda_{ee}$. At higher velocities, reducing the numerical drag and diffusion coefficients below that of real Coulomb collisions may require an even larger number of macroparticles.

While the situation may seem hopeless for schemes with a resolved Debye radius, if relaxation or partial thermalization due to real Coulomb collisions is an important process in the plasma which is to be simulated, the numerical thermalization can to some degree act as a substitute for real Coulomb collisions. However, it is important to compare the rate of this rapid thermalization to other relevant timescales to assess whether it will affect the results.

In the next section, we will provide several examples of exactly this analysis for a variety of low temperature plasma simulations in 2D PIC.

## 4. Example applications to 2D PIC simulations of low temperature plasma sources

### 4.1 Electron beam-generated plasma in a magnetic field

Electron beam-generated plasmas are of interest as a source of plasmas with large densities ($10^{15}$-$10^{17}$ $m^{-3}$) and low electron temperatures ($\lesssim 1\ eV$) suitable for materials processing applications. [30, 31, 2] This can produce a large flux of low energy ($< 5\ eV$) ions potentially useful for atomic layer etching or deposition.

Here we examine the effects of numerical thermalization on a plasma generated by a beam of electrons, such as that described by Rauf et al. [2] The explicit 2d3v open-source PIC code EDIPIC-2D [10] was used to simulate a 2 keV beam passing through the center of a 9 cm by 12 cm chamber with grounded conducting boundaries. A symmetry plane passes through the center of the beam, seen in Fig. 5A. The beam ionizes background neutral argon gas and an applied magnetic field runs parallel to the direction of the beam. The simulations were run for 800 μs, at which point a steady state was achieved.

Taking as an example a background gas pressure of 20 mTorr and an applied magnetic field strength of 100 Gauss, we can examine the EVDF within a region a few centimeters away from the beam. From Fig. 5B is clear to see that the electrons have a Maxwellian distribution function, though the EEDF is depleted somewhat at energies sufficient to overcome the large sheath potential. The density rapidly falls off with increasing distance from the beam (Fig. 5C), while the electron temperature decreases steadily and remains essentially constant along field lines (Fig. 5D).

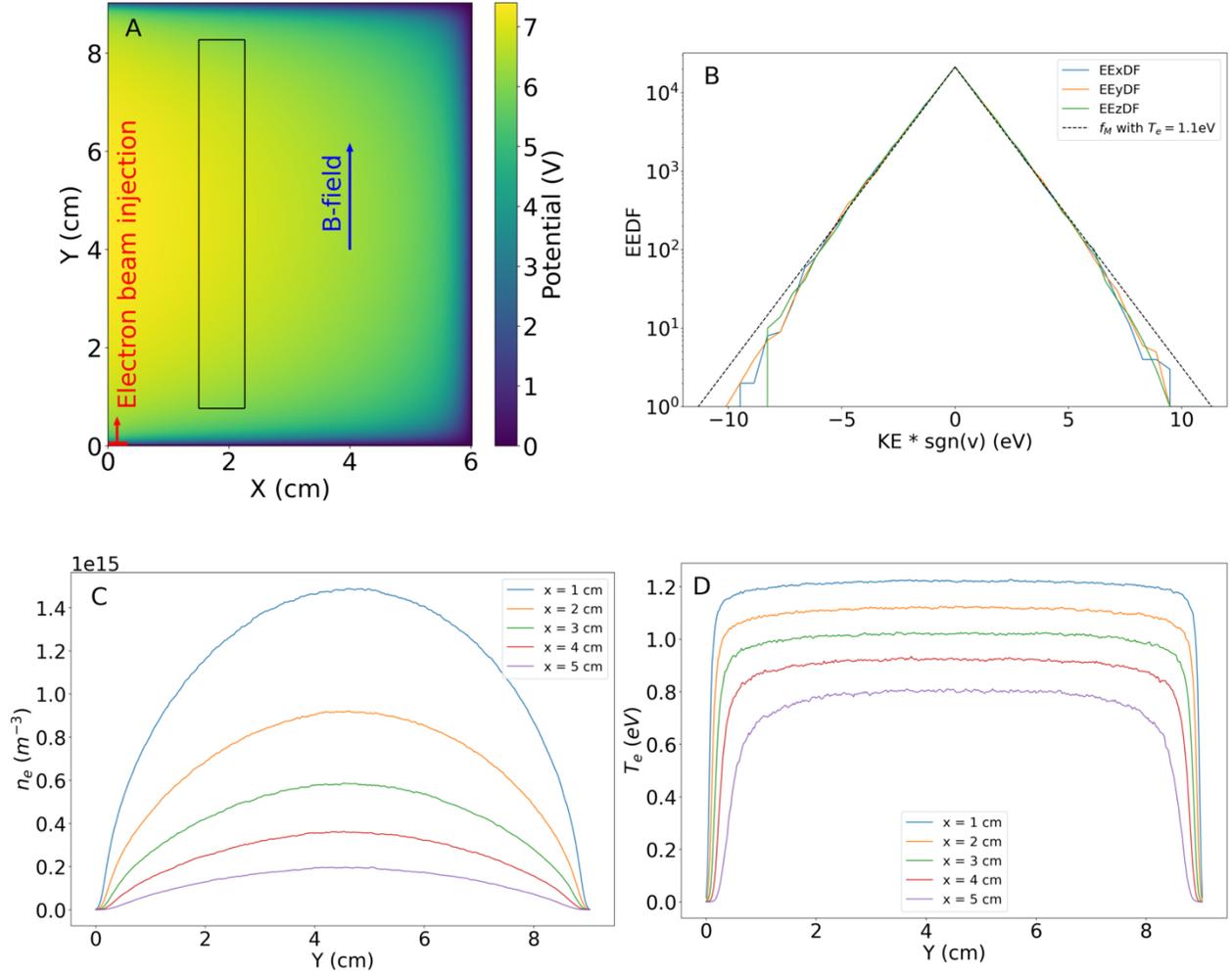

**Figure 5**: 2d3v PIC simulation of electron beam generated plasma. A) Colormap of electric potential in eV. The spatial region for EEDF averaging is outlined by a black rectangle. B) The one-dimensional EEDFs for electrons within the outlined spatial region, plotted against the kinetic energy in the corresponding direction times the sign of the velocity: $sgn(v_j)\frac{m_e}{2}v_j^2$ with $j \in \{x, y, z\}$. The black dashed line shows the theoretical EEDF of a Maxwellian distribution with a temperature of 1.1 eV. C) Electron density profiles along field lines at five different distances from the center of the beam. D) Electron temperature profiles along field lines at five different distances from the center of the beam (left boundary of Fig. 5A).

### 4.1.1 Timescale analysis: Electron beam-generated plasma

It should be emphasized that Coulomb collisions were not implemented in this simulation; the thermalization process is entirely numerical. To understand how it has impacted the results of the simulation, we should compare the timescale for numerical thermalization to the time required for electrons to diffuse away from the beam. These in turn should be compared to timescales associated with thermalization via real Coulomb collisions.

Based on the electron fluid velocity in the x-direction $\bar{v}_x$ observed in the simulations, we can estimate the time required for an electron to travel 2 cm away from the beam center along the x direction to be $t_x = 250$ µs. On the other hand, an electron with sufficient kinetic energy (say 8 eV) to immediately escape along the magnetic field lines to the upper or lower wall will do so in a time less than a few dozen nanoseconds ($t_y = 27\ ns$), assuming it doesn't experience an elastic collision with a neutral which redirects it. For electrons with a kinetic energy between 1 eV and 8 eV the electron-neutral elastic collision time ranges from $t_{e-Ar} = 6.75\ ns$ to $t_{e-Ar} = 180\ ns$. Here $t_{e-Ar} = 1/\nu_{e-Ar}$ is the inverse of the electron elastic collision rate, where $\nu_{e-Ar} = n_{Ar}\sigma_{el}(E)\sqrt{2E/m_e}$ depends on the background gas density $n_{Ar}$ and the collision cross section $\sigma_{el}(E)$ for an electron with kinetic energy $E$.

The numerical thermalization timescale falls between the two travel times. With an electron temperature of 1.3 eV and plasma density of $2 \times 10^{15} m^{-3}$, the electron Debye length near the beam is 0.19 mm and the plasma frequency is $2.5 \times 10^9\ s^{-1}$. The grid resolution is 0.21 mm and there are roughly 200 macroparticles per cell. Using Hockney's empirical estimate of the thermalization time, we would have $\tau_R^{num} \approx 1.1$ µs. Clearly, the thermalization occurs well before the electrons have diffused beyond the beam region, as confirmed in Fig. 4. However, it occurs slowly enough that the fast electrons with a kinetic energy larger than the sheath potential will almost certainly have time to escape before they are significantly slowed down.

How does this numerical thermalization compare to real thermalization timescales via Coulomb collisions? Taking as our estimate for the Coulomb thermalization timescale the electron-electron slowing down time for a test particle travelling at the electron thermal velocity in a Maxwellian plasma, we have $\tau_R^{real} \approx 4.5$ μs. While this is larger than the numerical thermalization time, it is still quite small compared to the time required for the electrons to diffuse a few centimeters away from the beam. It is also large compared to the time required for fast electrons to escape to the upper and lower walls along magnetic field lines. This analysis can be summed up by the following relation:

$$t_x > \tau_R^{real} > \tau_R^{num} > t_{e-Ar} > t_y \tag{17}$$

Because the thermalization caused by numerical noise happens on a timescale that can be ordered with respect to travel/escape timescales in a manner similar to the real thermalization time due to Coulomb collisions, we can conclude that the numerical thermalization in the beam region will more or less reproduce the effects of real Coulomb collisions, though it occurs at a more rapid rate.

### 4.1.2   Possible effects of numerical noise: Electron beam-generated plasma

There are some complications when we consider the flux of electrons to the wall in regions far from the beam. In these areas, the cold trapped electrons contained by the large sheath potentials must escape to the wall by diffusing outwards in velocity space, as there is no other mechanism for them to gain kinetic energy. The fact that the EEDF is depleted above the energy required to overcome the sheath potential indicates that the flux to the wall is limited by the flux of electrons into the untrapped region of phase space. (This is analogous to the loss cone of a magnetic mirror with an additional confining electric potential; here the region is bounded by two planes in velocity space since the magnetic field strength is constant.) The only mechanism for electrons to gain energy is through parallel diffusion in velocity space – which occurs more rapidly due to numerical noise than the analogous process caused by real Coulomb collisions.

If we were to hypothetically increase the number of macroparticles per cell and thus decrease the rate of numerical collisions, we may at one point achieve a rate of refilling the untrapped region of phase space comparable to the rate caused by real Coulomb collisions. However, without implementing Coulomb collisions into the code, there would be no indication that this state had been achieved – and further increasing the number of particles per cell would cause the outward diffusion of electrons in velocity space to occur ever more slowly.

**4.2 Hollow cathode plasma**

Here we analyze the impact of numerical thermalization on a 2D PIC simulation of a hollow cathode discharge in an argon plasma, also performed using EDIPIC-2D [10]. Electrons are emitted by the cathode with 50 A of current into a narrow channel 1 cm in diameter and 10 cm in length, ionizing a background argon gas at a pressure of 93 mTorr. Electrons produced via ionization expand out of the channel and into the larger bulk plasma, accelerating towards the anode above the channel opening, which is held at 30 V. In this simulation, the permittivity of free space was scaled by a factor of 1059, reducing the plasma frequency by a factor of about 32.5. The electron velocity distribution function within the channel after a total simulation time of 200 μs is shown in Fig. 6B, averaged over the channel area shown in Fig. 6A.

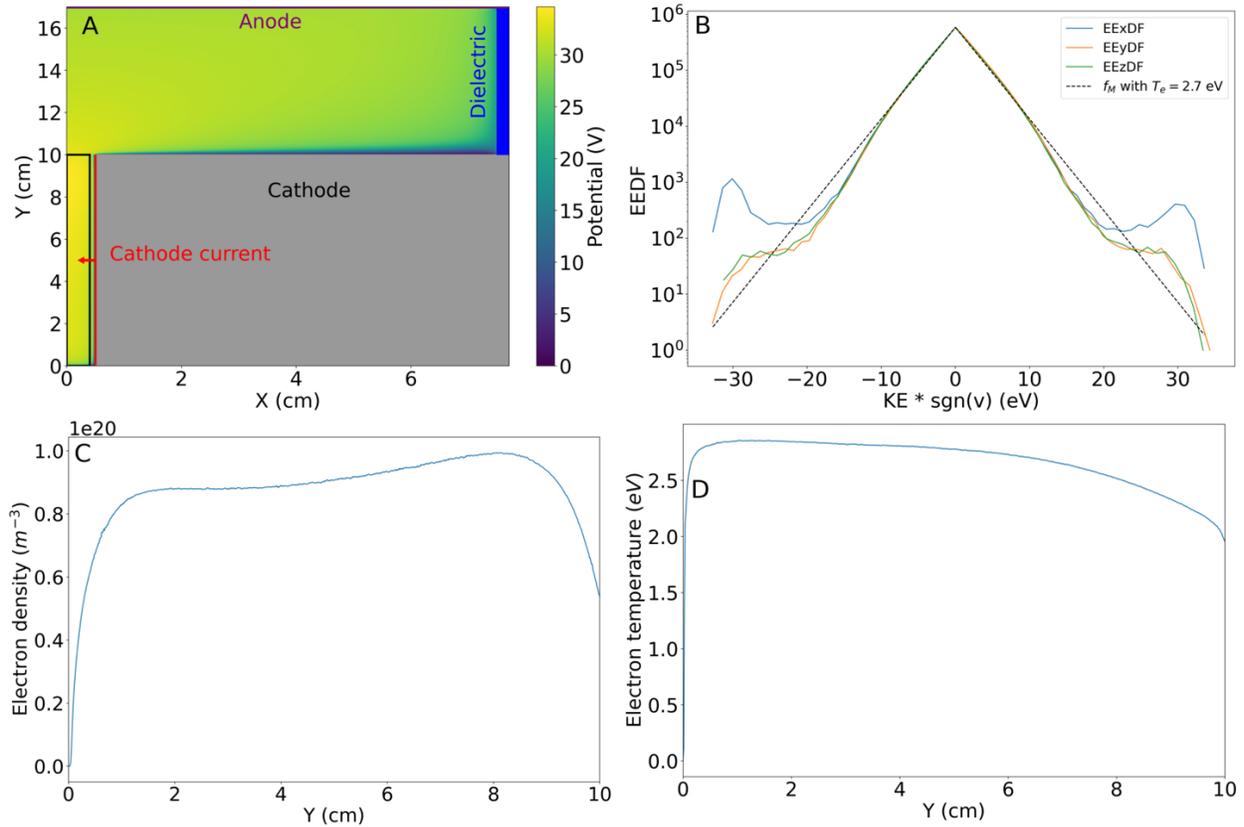

**Figure 6**: 2d3v PIC simulation of a hollow cathode plasma. A) Colormap of electric potential in eV. The spatial region for EEDF averaging is outlined by a black rectangle. The left boundary is a reflective symmetry axis. B) The EEDF for electrons within the outlined spatial region. The black dashed line shows the theoretical EEDF of a Maxwellian distribution with a temperature of 2.7 eV. C) x-averaged electron density along the length of the channel, within the spatial region outlined by the black rectangle in Fig. 6A. D) x-averaged electron temperature (mean kinetic energy) along the length of the channel within the spatial region outlined by the black rectangle in Fig. 6A.

Within the channel a population of cold electrons at energies less than 10 eV are well described by a Maxwellian distribution with a temperature of 2.7 eV. Above this energy, the EEDF falls off faster than a Maxwellian until the peaks representing a population of fast electrons injected from the cathode is visible at roughly 30 eV. These peaks are especially pronounced in the one-dimensional EEDF obtained from the velocity in the x-direction (see Fig. 6B) as the strong electric field from the sheath potential at the cathode wall accelerates injected electrons across the channel. However, the one-dimensional EEDFs in the y- and z-directions are also enhanced at high energies due to elastic scattering of the accelerated electrons.

### 4.2.1 Timescale analysis: Hollow cathode plasma

Based on the electron fluid velocity in the y-direction $\bar{v}_y$ observed in the simulations, we can estimate the time required for an electron to travel the length of the channel (10 cm) to be $t_y = 13 \ \mu s$. A negligible number of electrons return to the cathode wall, as they would require nearly 35 eV of kinetic energy in the x-direction. Such an electron unimpeded by collisions would travel to the wall in $t_x = 1.4 \ ns$. The elastic collision timescale for a thermal electron at 2.7 eV in this simulation is $t_{e-Ar} = 9 \ ns$, while the same for 35 eV electron is only $t_{e-Ar} = 1.5 \ ns$.

With an electron density of roughly $10^{20} \ m^{-3}$ and a temperature of 2.7 eV in the channel region, the plasma frequency is about $5.6 \times 10^{11} \ s^{-1}$ and the Debye length is 1.2 µm. However, due to the scaling of the permittivity of free space, the "simulated" plasma frequency and Debye length are $1.7 \times 10^{10} \ s^{-1}$ and 38 µm, respectively. Based on a macroparticle density of $n_{mac} = 3.12 \times 10^{10} \ m^{-2}$ and a cell width of 125 µm, this yields a thermalization time estimate of $\tau_R^{num} \approx 200 \ ns$. In comparison, the thermalization timescale from (hypothetical, not implemented) Coulomb collisions (estimated as in the previous section) is $\tau_R^{real} \approx 1.6 \ ns$. We can once again order these timescales:

$$t_y > \tau_R^{num} > t_{e-Ar} \sim t_x \sim \tau_R^{real} \tag{18}$$

In this case, the scaling of the permittivity of free space has actually reduced the rate of thermalization below that caused by real Coulomb collisions. Both simulated and real thermalization rates occur faster than physical transport of electrons along the channel.

### 4.2.2 Inclusion of real electron-electron collisions

As shown in the previous section, scaling the permittivity of free space reduces numerical collisions below the rate of real electron-electron collisions. Thus, simulations which include Coulomb collisions will no longer be overwhelmed by numerical thermalization and depend wholly on real processes. This can be seen in the channel EEDF of the following hollow cathode simulation, using a similar geometry to the setup described previously, identical background gas pressure, and with an emitted current density at the cathode of $2.06\ kA \cdot m^2$. Thermalization here is driven by implemented electron-electron Coulomb collisions using the Nanbu method [32], and numerical collisions are reduced through scaling of the permittivity of free space. The electron-electron collision frequency in this case has been made faster than the estimated numerical thermalization rate by a factor of 20.

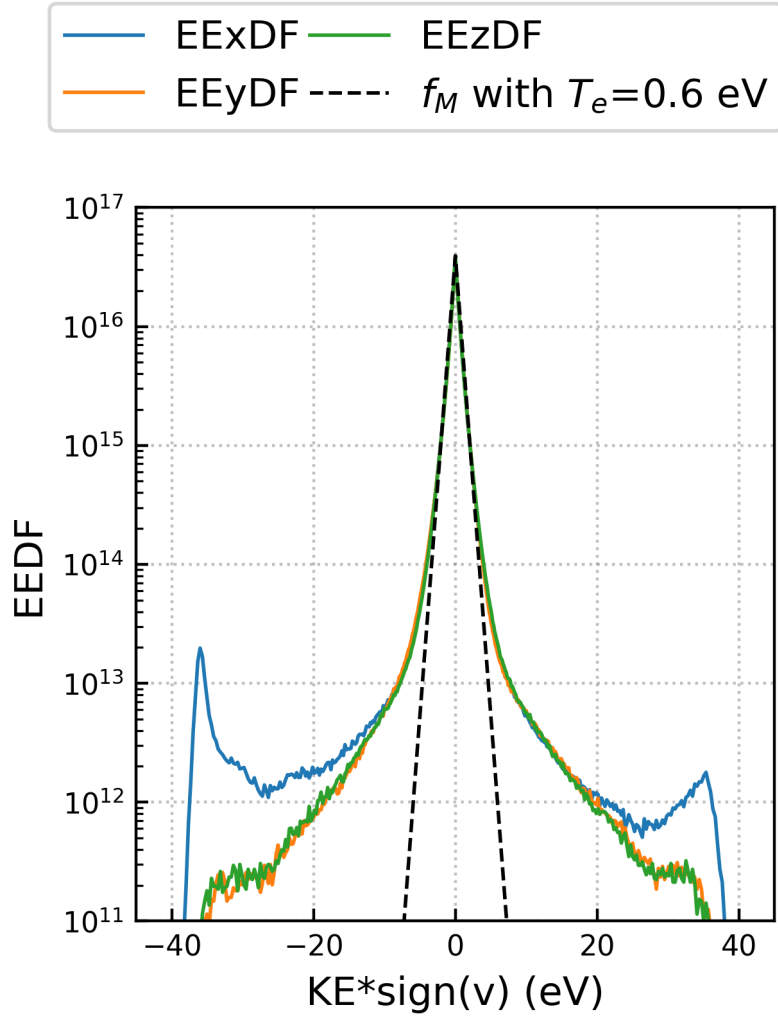

**Figure 7:** The one-dimensional EEDFs for electrons within the channel of the hollow cathode. The black dashed line shows the theoretical EEDF of a Maxwellian distribution with a temperature of 0.6 eV.

This EEDF in Fig. 7 is clearly non-Maxwellian; the high energy electrons accelerated across the sheath from the cathode maintain large peaks in the wings, even more pronounced than in figure 6B. Though the cold electrons have begun to thermalize, we can attribute this to the actual impact of Coulomb collisions rather than the numerical thermalization which would otherwise dominate the process if permittivity of free space had not been scaled to mitigate it.

## 4.3 Inductively-coupled plasma

The following 2D PIC simulation is that of an inductively coupled plasma operating in argon at 5 mTorr background gas pressure. The following electromagnetic simulation was performed using a direct implicit scheme and Darwin algorithm in an adaptation of the EDIPIC-2D code. The antennae within the dielectric carry a 120 A current that oscillates at a frequency of 2 MHz and the non-dielectric boundaries of the simulation space are grounded metal walls. The simulation ran for 100 μs before the snapshots depicted in Fig. 8 were taken.

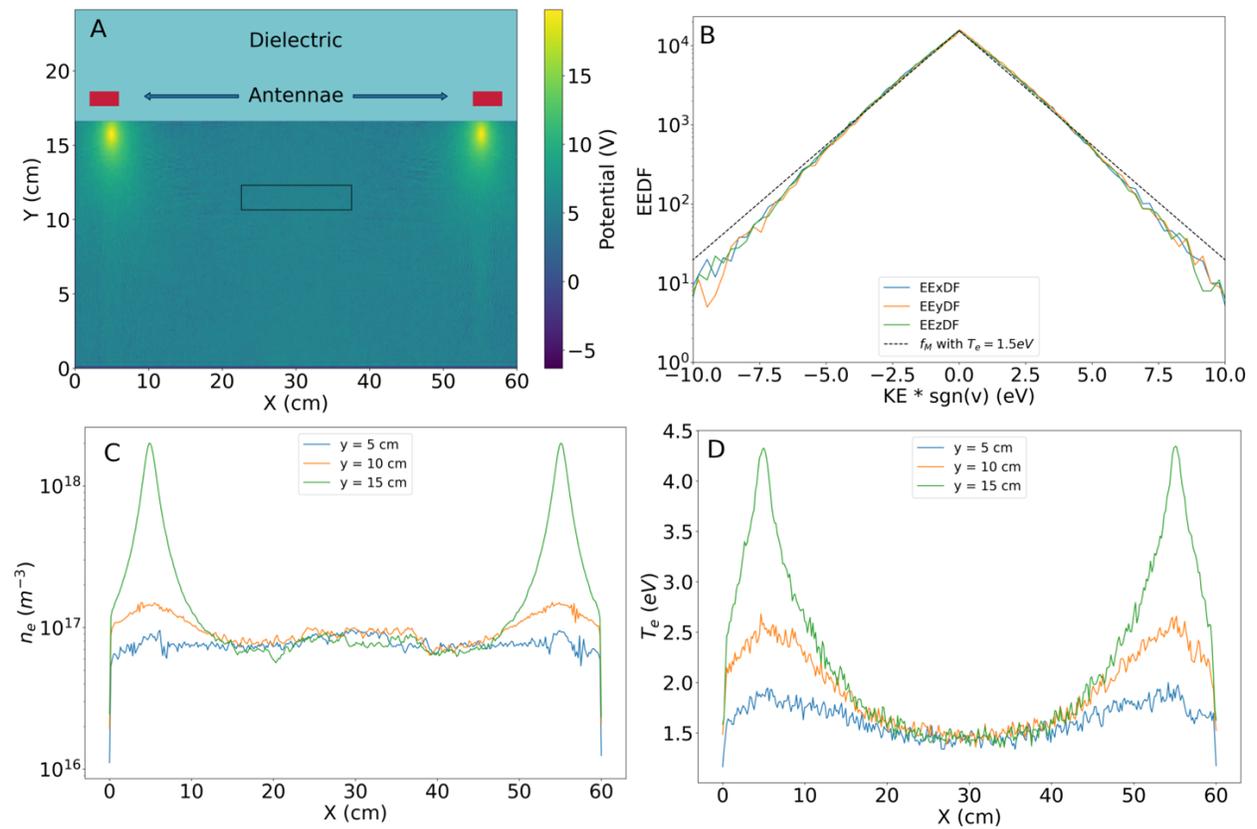

**Figure 8**: 2d3v PIC simulation of an inductively coupled plasma. A) Colormap of electric potential in eV. Note that the bounding conductive walls are grounded, and the sheath is not resolved. The spatial region for EEDF averaging is outlined by a black rectangle. B) The EEDF for electrons within the outlined spatial region. The black dashed line shows the theoretical EEDF of a Maxwellian distribution with a temperature of 1.5 eV.

C) Electron density profiles at three different distances from the dielectric. D) Electron temperature profiles at three different distances from the dielectric.

### 4.3.1 Timescale analysis: Inductively-coupled plasma

A thermal 1.5 eV electron would take $t_{x,y} \approx 210\ ns$ to travel 15 cm without collisions, while the electron-neutral elastic collision time for such an electron is $t_{e-Ar} = 410\ ns$. With an electron density of $8.7 \times 10^{16}\ m^{-3}$ and electron temperature of about 1.5 eV, the plasma frequency within the outlined region in Fig. 8 is $1.7 \times 10^{10}\ s^{-1}$ and the Debye length is 31 μm. Considering a cell size of 1.66 mm and 1000 particles per cell, this yields a thermalization time based on Hockney's estimate of $\tau_R \approx 380$ ns. Thermalization due to real Coulomb collisions is estimated to occur on a timescale of $\tau_R \approx 590$ ns.

$$\tau_R^{real} > \tau_R^{num} \sim t_{e-Ar} > t_{x,y} \tag{19}$$

The numerical thermalization thus reasonably approximates the effects of real Coulomb collisions, and for trapped cold electrons will occur roughly on the same timescale as it takes for electrons to travel the length of the system. As with the electron beam simulation, complications arise when we consider the fast-moving electrons in the depleted wings of the distribution at kinetic energies sufficient to overcome the sheath potential. For those electrons, numerical diffusion in velocity space is likely occurring faster than it would be if it were caused by real Coulomb collisions.

### 4.4 Capacitively-coupled plasma

Capacitively coupled plasmas are frequently used in materials processing applications. In the following 2d3v simulation of a capacitively coupled argon plasma with a background gas pressure of 10 mTorr, we examine the potential and EEDF distribution averaged over the last 1.5 µs of a 295 µs simulation time. (See Fig. 9.) A powered electrode at the bottom of the simulation space (colored grey in Fig. 9A) is pulsed at a frequency of 13.56 MHz and is separated from the grounded outer walls by a dielectric (in blue). The amplitude of the applied voltage is 100 V. The simulation was run using the Low-Temperature Particle in cell code (LTP-PIC), a massively parallel, GPU accelerated particle-in-cell code specifically designed for modeling low-temperature plasma devices. [29] The code has been benchmarked in the collisionless [33] and collisional regime. [3]

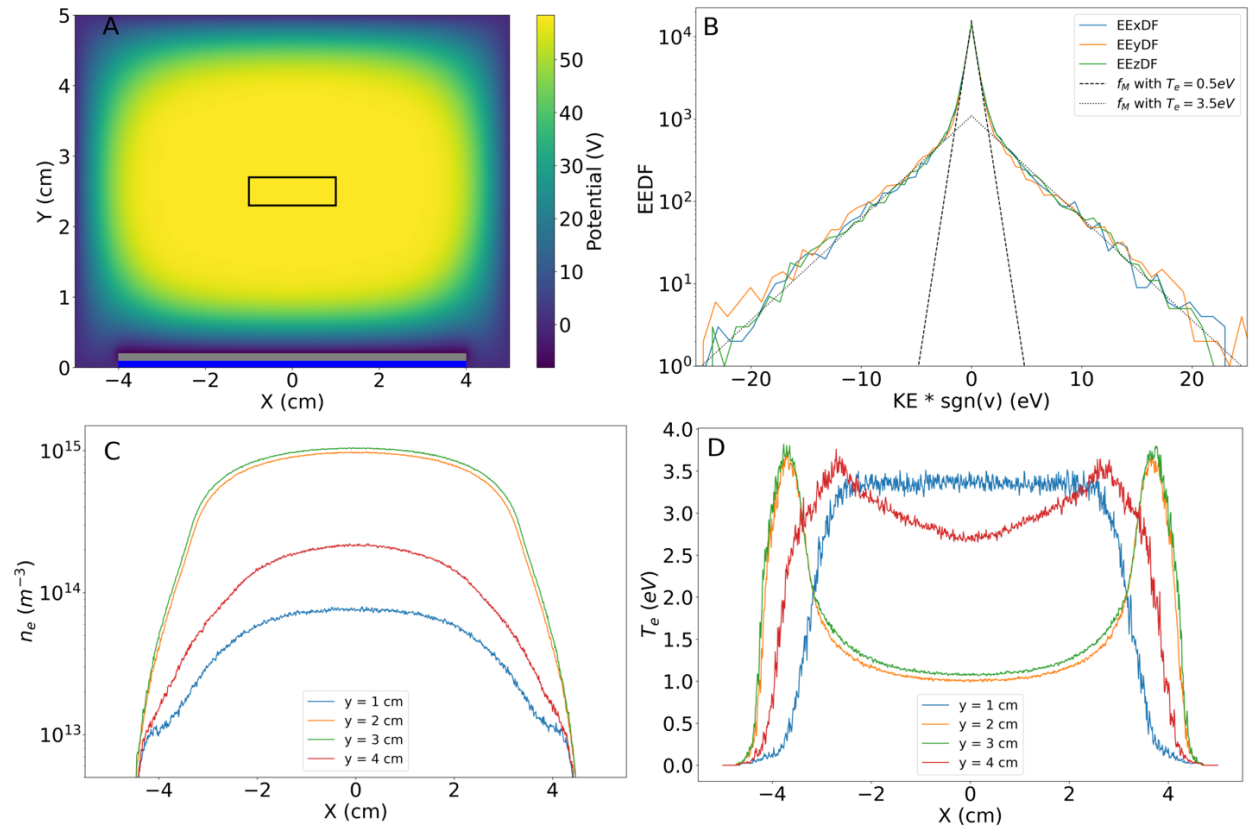

**Figure 9**: 2d3v simulation of a capacitively coupled plasma, averaged over 1.5 µs. A) Colormap of time-averaged electric potential in eV. The bounding walls are grounded, the powered electrode is in grey, and the dielectric is in blue. The spatial region for EEDF averaging is outlined by a black rectangle. B) The EEDF for electrons within the outlined spatial region. The black dashed line shows the theoretical EEDF of a Maxwellian

distribution with a temperature of 0.5 eV, while the black dotted line shows an EEDF with a temperature of 3.5 eV. C) Electron density profiles at four different distances from the powered electrode. D) Electron temperature profiles at four different distances from the powered electrode.

The time-averaged EEDF from the spatial region outlined in black in Fig. 9A shows a characteristic bi-Maxwellian character of low pressure CCP discharges. *[34]* (See Fig. 7B.) Cold electrons are trapped by the relatively flat potential, while the hotter population experiences heating through interactions with the fluctuating sheath near the electrode. Unlike the EEDFs described by Godyak et al, *[34]* the EEDF produced in this simulation does not sharply drop near the excitation energy of argon at 11.5 eV. The densities and temperatures of the high temperature and low temperature electron populations can be obtained from the best fit in the high energy and low energy regions as in He et al. *[35]*.

$$f(E) = f_{low}(E) + f_{high}(E) = Ae^{-E/T_{e\_low}} + Be^{-E/T_{e\_high}} \tag{20}$$

$T_{e\_high}$ can be immediately extracted from the fit to the distribution at higher energies, where $e^{-E/T_{e\_high}} \gg e^{-E/T_{e\_low}}$, while the low temperature population can then be defined as $f_{low}(E) = f(E) - f_{high}(E)$ and the relative density and temperature extracted accordingly. Thus, the distribution shown in Fig. 7B yields a high temperature electron population with $T_{e\_high} = 3.5\ eV$ and a low temperature population with $T_{e\_low} = 0.6\ eV$, the former population having a number density roughly 80 times that of the latter.

### 4.4.1 Timescale analysis: Capacitively-coupled plasma

The numerical thermalization clearly does not occur so quickly that electrons have been thermalized to a uniform temperature. Electrons with an energy of 3.5 eV, unimpeded by collisions, will travel 5 cm in $t_{x,y} = 45$ ns. The electron-neutral elastic collision time for an electron with 3.5 eV of kinetic energy is about $t_{e-Ar} = 56\ ns$. One oscillation of the applied voltage occurs over a time period of about $\tau_{RF} = 460\ ns$.

Considering the simulated time-averaged electron density in the region of interest is $1.1 \times 10^{15}\ m^{-3}$, the plasma frequency is $1.9 \times 10^9\ s^{-1}$. Using a temperature of 3.5 eV, the Debye length is $0.42\ mm$. With a grid resolution of $0.1\ mm$ and a macroparticle weighting of fifty thousand simulated electrons per macroparticle, we might estimate the thermalization time as $\tau_R^{num} \approx 1.4\ \mu s$. In contrast, the timescale of thermalization due to real Coulomb collisions, estimated as in previous sections, is $\tau_R^{real} \approx 130\ \mu s$. This results in the ordering:

$$\tau_R^{real} > \tau_R^{num} > \tau_{RF} > t_{e-Ar} \sim t_{x,y} \qquad (21)$$

Numerical thermalization occurs two orders of magnitude faster than real thermalization. Though the bi-Maxwellian EEDF is produced, it is possible that the more rapid numerical thermalization has smoothed out the very high energy portion of the EEDF so that there is no rapid drop off beyond the electron excitation energy at 11.5 eV.

It has been observed by Vass et al that for some (one-dimensional) PIC simulations of low pressure RF discharges, regardless of how many particles per cell are used, the simulations will not converge unless Coulomb collisions are incorporated into the code [4]. (Notably, the dependence of the EEDF at low energies on electron-electron Coulomb collisions in RF discharges has been discussed analytically by Kaganovich and Tsendin [36], and the generation of cold trapped electrons in this scenario is sensitive to the rate of Coulomb collisions [37].) As discussed for beam electrons, this is likely due to the fact that the diffusion of electrons in velocity space will at some point depend on numerical collisions unless the effective numerical collision rate is made smaller than that of implemented electron-electron Coulomb collisions.

## 5. Conclusions

We have calculated the drag and diffusion coefficients for the Fokker-Planck form of the numerical collision operator for 2D electrostatic PIC simulations employing the commonly used CIC particle weighting scheme. It is explained that these drag and diffusion coefficients can be used to assess numerical thermalization timescales, which generally depend on the number of macroparticles per Debye volume. Via simple scaling it is easily demonstrated that the number of macroparticles required to suppress the rate of numerical thermalization below the rate of thermalization caused by Coulomb collisions is quite large, necessitating that each macroparticle represent only a handful of real particles. This requirement may be achievable if the plasma which is being simulated has a sufficiently small plasma parameter.

Grid spacings larger than the Debye length can reduce the numerical drag and diffusion coefficients by increasing the size of the macroparticle shape functions, but these reductions will be negligible for electrons travelling at many times the electron thermal velocity, $v_{Te}$. If the simulated permittivity of free space is increased by some large factor, then the effective simulated plasma frequency and thus the thermal relaxation rate will also be reduced.

The slowing down timescale associated with the drag coefficient is shown to qualitatively match the empirical formula developed by Hockney in 1970 [8]. While Hockney recommended that the thermalization timescale should be used to set an upper limit on the length of the simulation, this need not be the case if one can ensure that the timescale of numerical thermalization can be ordered with other timescales that affect the EVDF in the same way that the real thermalization time due to Coulomb collisions would be ordered. Practical examples of this thermalization timescale investigation have been given for a variety of low temperature plasma simulations.

One feature of note for both the electron beam generated plasma and ICP simulations is that the EEDF is depleted at energies larger than the sheath potential. In order for cold electrons to escape to the wall, they must be able to diffuse outwards in velocity-space. Without implemented Coulomb collisions, the only mechanism by which cold electrons can gain energy is through numerical parallel diffusion in velocity space.

For electrons with a kinetic energy just below that of the sheath potential, this numerical parallel diffusion in two dimensional simulations often happens much more rapidly than an analogous process caused by Coulomb collisions. Because of this, the loss region in velocity space is more rapidly refilled, and the electron wall fluxes are likely higher than they otherwise would be if the parallel diffusion were caused by Coulomb collisions between real particles. This is something to bear in mind when using PIC codes to simulate plasmas which contain populations of trapped cold electrons.

**Acknowledgements:**

The research described in this paper was supported by the US Department of Energy under Laboratory Directed Research & Development (LDRD) program at Princeton Plasma Physics Laboratory (PPPL) and the PPPL CRADA agreement with Applied Materials entitled

**Appendix A:**

Here we give an expression for the numerical collision operator for species $\alpha$ in an infinite homogeneous PIC plasma with a 2D cartesian grid, as derived by Touati et al [13]. We assume that the time step is sufficiently small in comparison to the electron plasma period that aliasing effects due to the timestep can be neglected, that the grid spacing is equal in both spatial directions, and that a CIC scheme has been used for particle shape functions.

$$\left(\frac{\partial f_\alpha}{\partial t}\right)_c = \sum_\beta \frac{n_\beta}{m_\alpha} \frac{\partial}{\partial v} \cdot \int d\boldsymbol{v}' \, \boldsymbol{Q}_{\alpha\beta}(\boldsymbol{v},\boldsymbol{v}') \cdot \left(\frac{\delta N_\beta}{m_\alpha}\frac{\partial}{\partial v} - \frac{\delta N_\alpha}{m_\beta}\frac{\partial}{\partial v'}\right) f_\alpha(\boldsymbol{v}) f_\beta(\boldsymbol{v}') \quad (A.1)$$

$$\boldsymbol{Q}_{\alpha\beta}(\boldsymbol{v},\boldsymbol{v}') \equiv \int \frac{d\boldsymbol{k}}{(2\pi)^2} \boldsymbol{K} \otimes \boldsymbol{K} \sum_p \left|\frac{S(\boldsymbol{k})S(\boldsymbol{k}_p)q_\alpha q_\beta}{\varepsilon_0 \epsilon(\boldsymbol{k}\cdot\boldsymbol{v},\boldsymbol{k})K^2}\right|^2 \pi\delta(\boldsymbol{k}\cdot\boldsymbol{v} - \boldsymbol{k}_p\cdot\boldsymbol{v}') \quad (A.2)$$

$$\epsilon(\omega,\boldsymbol{k}) \equiv 1 + \sum_\gamma \frac{\omega_{p\gamma}^2}{K^2} \sum_p S(\boldsymbol{k}_p)^2 \int d\boldsymbol{v}'' \frac{\partial f_\gamma}{\partial v''} \cdot \frac{\boldsymbol{K}}{\omega - \boldsymbol{k}_p \cdot \boldsymbol{v}''} \quad (A.3)$$

$$\boldsymbol{k}_p = \boldsymbol{k} - p\boldsymbol{k}_g, \qquad k_g = \frac{2\pi}{\Delta_x}, \qquad S(\boldsymbol{k}) = \text{sinc}(k_x\Delta_x/2)^2 \, \text{sinc}(k_y\Delta_x/2)^2 \quad (A.4)$$

This collision operator is written such that particle density is factored out of the velocity distribution functions, so $1 = \int d\boldsymbol{v}\, f_\alpha(\boldsymbol{v})$. The macroparticle weighting factor, $\delta N_\alpha$, scales the macroparticle densities ($n_{\alpha,mac}$), masses ($M_\alpha$), and charges ($Q_\alpha$) to that of real particles such that $\delta N_\alpha = n_\alpha/n_{\alpha,mac} = M_\alpha/m_\alpha = Q_\alpha/q_\alpha$.

Vector quantities have two directions; volume integrals are defined as one would expect; for example, $d\boldsymbol{k}$ represents $dk_x dk_y$ and similarly for $d\boldsymbol{v}'$. Aliasing effects are accounted for through the sums over all integers $\boldsymbol{p} = (p_x, p_y)$ which couple various aliased wavelengths and frequencies, and $\Delta_x$ is the grid resolution.

The vector $\boldsymbol{K}$ depends on the specific Maxwell solver used. Touati et al give the following expressions for its $j^{\text{th}}$ component when a spectral Maxwell solver or alternately the second order finite-difference time-domain (FDTD) method of Yee [38] coupled with the charge conserving scheme of Villasenor and Buneman [39] is implemented:

$$K_j = \begin{cases} k_j & \text{for the spectral Maxwell solver} \\ k_j \operatorname{sinc}(k_j \Delta_j/2) & \text{for the 2nd order FDTD Maxwell solver} \end{cases}$$

We assume quasineutrality and equal macroparticle weighting of the electrons and ions limiting our consideration to one singly-charged ion species: $n_\beta = n$ and $\delta N_\beta = \delta N = n/n_{mac}$. We can then write:

$$\left(\frac{\partial f_\alpha}{\partial t}\right)_c = -\frac{\partial}{\partial \boldsymbol{v}} \cdot [\boldsymbol{A}_{\alpha 1}(\boldsymbol{v}) f_\alpha(\boldsymbol{v})] + \frac{1}{2}\frac{\partial}{\partial \boldsymbol{v}} \cdot \left[\boldsymbol{D}_\alpha(\boldsymbol{v}) \cdot \frac{\partial f_\alpha(\boldsymbol{v})}{\partial \boldsymbol{v}}\right] \qquad (A.5)$$

$$\boldsymbol{A}_{\alpha 1}(\boldsymbol{v}) = \sum_\beta \frac{n^2}{n_{mac} m_\alpha m_\beta} \int d\boldsymbol{v}' \, \boldsymbol{Q}_{\alpha\beta}(\boldsymbol{v}, \boldsymbol{v}') \cdot \frac{\partial f_\beta(\boldsymbol{v}')}{\partial \boldsymbol{v}'} \tag{A.6}$$

$$\boldsymbol{D}_\alpha(\boldsymbol{v}) = \sum_\beta \frac{2n^2}{n_{mac} m_\alpha^2} \int d\boldsymbol{v}' \, \boldsymbol{Q}_{\alpha\beta}(\boldsymbol{v}, \boldsymbol{v}') f_\beta(\boldsymbol{v}') \tag{A.7}$$

This is a common way of writing the collision operator frequently referred to as the Fokker-Planck equation. It allows for better comparison with previous results regarding drag and diffusion in two dimensional plasmas. [18, 17, 19] However, while $\boldsymbol{A}_{\alpha 1}(\boldsymbol{v})$ is the largest portion of the electron-electron drag term at small velocities (reflecting polarization drag), the dynamical friction drag term has been neglected:

$$\boldsymbol{A}_{\alpha 2}(\boldsymbol{v}) = \frac{1}{2} \frac{\partial}{\partial \boldsymbol{v}} \cdot \boldsymbol{D}_\alpha(\boldsymbol{v}), \qquad \boldsymbol{A}_\alpha(\boldsymbol{v}) = \boldsymbol{A}_{\alpha 1}(\boldsymbol{v}) + \boldsymbol{A}_{\alpha 2}(\boldsymbol{v}) \tag{A.8}$$

$$\boldsymbol{A}_\alpha(\boldsymbol{v}) = \sum_\beta \frac{n^2}{n_{mac} m_\alpha^2} \left(1 + \frac{m_\alpha}{m_\beta}\right) \int d\boldsymbol{v}' \, \boldsymbol{Q}_{\alpha\beta}(\boldsymbol{v}, \boldsymbol{v}') \cdot \frac{\partial f_\beta(\boldsymbol{v}')}{\partial \boldsymbol{v}'} \tag{A.9}$$

When dynamical friction is incorporated, we can write the proper Fokker-Planck form with $\boldsymbol{A}_\alpha(\boldsymbol{v})$ representing the drag coefficient.

$$\left(\frac{\partial f_\alpha}{\partial t}\right)_c = -\frac{\partial}{\partial \boldsymbol{v}} \cdot [\boldsymbol{A}_\alpha(\boldsymbol{v}) f_\alpha(\boldsymbol{v})] + \frac{1}{2} \frac{\partial}{\partial \boldsymbol{v}} \frac{\partial}{\partial \boldsymbol{v}} : [\boldsymbol{D}_\alpha(\boldsymbol{v}) f_\alpha(\boldsymbol{v})] \tag{A.10}$$

We neglect the effects of discrete time steps, assuming that the timestep is sufficiently small to make a negligible impact. We then evaluate the drag and diffusion coefficients for a test electron in a background of Maxwellian electrons and ions with distributions: $f_\beta(v) = \frac{1}{\pi v_{T\beta}^2} e^{-v^2/v_{T\beta}^2}$.

Writing $\mathbf{A}_{e1}(v) = \sum_\beta \mathbf{A}_1^{e/\beta}(v)$ and $\mathbf{D}_e(v) = \sum_\beta \mathbf{D}^{e/\beta}(v)$, we have

$$\mathbf{A}^{e/\beta}(v) = -\frac{\omega_{pe}^4}{2\pi^{3/2} n_{mac}} \left(1 + \frac{m_e}{m_\beta}\right) \frac{1}{v_{T\beta}^3} \int d\mathbf{k} \sum_p \left|\frac{S(\mathbf{k})S(\mathbf{k}_p)}{\epsilon(\mathbf{k}\cdot\mathbf{v},\mathbf{k})K^2}\right|^2 \frac{\mathbf{K}(\mathbf{K}\cdot\mathbf{k}_p)(\mathbf{k}\cdot\mathbf{v})}{k_p^3} e^{-(\mathbf{k}\cdot\mathbf{v})^2/k_p^2 v_{T\beta}^2} \qquad (A.11)$$

$$\mathbf{D}^{e/\beta}(v) = \frac{\omega_{pe}^4}{2\pi^{3/2} n_{mac}} \frac{1}{v_{T\beta}} \int d\mathbf{k} \sum_p \left|\frac{S(\mathbf{k})S(\mathbf{k}_p)}{\epsilon(\mathbf{k}\cdot\mathbf{v},\mathbf{k})K^2}\right|^2 \frac{\mathbf{K}\otimes\mathbf{K}}{k_p} e^{-(\mathbf{k}\cdot\mathbf{v})^2/k_p^2 v_{T\beta}^2} \qquad (A.12)$$

with plasma frequency $\omega_{pe}^2 = nq^2/m_e \varepsilon_0$.

Scaling velocities by the thermal electron velocity ($\mathbf{v} = \mathbf{V} v_{Te}$ and $v_{T\beta} = R_{v\beta} v_{Te}$) and wavenumber quantities by the electron Debye length ($\mathbf{k} = \mathbfit{k}/\lambda_{De}$, $\mathbf{k}_p = \mathbfit{k}_p/\lambda_{De}$, $\mathbf{K} = \mathbfit{K}/\lambda_{De}$ with $\lambda_{De} = \sqrt{k_B T_e \varepsilon_0/nq^2}$) we arrive at expressions for the drag and diffusion with dimensionless integrals.

$$\mathbf{A}^{e/\beta}(v) = -\frac{\omega_{pe} v_{Te}}{2^{5/2}\pi^{3/2} N_{D,mac}} \left(1 + \frac{m_e}{m_\beta}\right) \frac{1}{R_{v\beta}^3} \int d\mathbfit{k} \sum_p \left|\frac{S(\mathbfit{k})S(\mathbfit{k}_p)}{\epsilon(\mathbfit{k}\cdot\mathbf{V},\mathbfit{k})\mathbfit{K}^2}\right|^2 \frac{\mathbfit{K}(\mathbfit{K}\cdot\mathbfit{k}_p)(\mathbfit{k}\cdot\mathbf{V})}{k_p^3} e^{-(\mathbfit{k}\cdot\mathbf{V})^2/k_p^2 R_{v\beta}^2} \qquad (A.13)$$

$$\mathbf{D}^{e/\beta}(v) = \frac{\omega_{pe} v_{Te}^2}{2^{5/2}\pi^{3/2} N_{D,mac}} \frac{1}{R_{v\beta}} \int d\mathbfit{k} \sum_p \left|\frac{S(\mathbfit{k})S(\mathbfit{k}_p)}{\epsilon(\mathbfit{k}\cdot\mathbf{V},\mathbfit{k})\mathbfit{K}^2}\right|^2 \frac{\mathbfit{K}\otimes\mathbfit{K}}{k_p} e^{-(\mathbfit{k}\cdot\mathbf{V})^2/k_p^2 R_{v\beta}^2} \qquad (A.14)$$

The numerical dielectric function evaluated for a quasineutral plasma with Maxwellian electrons and singly charged ions at temperatures $T_e$ and $T_i$, respectively, can be written:

$$\mathcal{D}(\boldsymbol{k} \cdot \boldsymbol{V}, \boldsymbol{k}) \equiv 1 - \sum_p \frac{S(\boldsymbol{k}_p)^2 (\mathcal{K} \cdot \boldsymbol{k}_p)}{2\mathcal{K}^2 k_p^2} \left[ Z'\left(\frac{\boldsymbol{k} \cdot \boldsymbol{V}}{k_p}\right) + \frac{T_e}{T_i} Z'\left(\frac{\boldsymbol{k} \cdot \boldsymbol{V}}{k_p R_{vi}}\right) \right] \quad \text{(A.15)}$$

Here $Z'(x)$ is the derivative of the plasma dispersion function evaluated at $x$, which was calculated using PlasmaPy. [40] We chose to use this full Balescu-Lenard collision operator analogue rather than simplifying the dielectric function to obtain a Landau collision operator analogue for the numerical collision operator based on the findings of Okuda and Birdsall who demonstrated that, unlike with point charges in three dimensions, the full collision operator is not well approximated by a Landau collision operator analogue in a reduced number of dimensions or with particles of finite width [17].

The drag and diffusion coefficients were found to have little dependence on the orientation of the test particle velocity with respect to the grid, and so for most of the following calculations $\boldsymbol{V}$ was chosen to point along the $\boldsymbol{k}_x$ axis. This would clearly not be the case for unequal grid spacing, in which case we would recommend using the smaller spacing size to obtain a larger estimate of the drag and diffusion coefficients scaled by macroparticle density. The number of particles per cell can then be adjusted to account for the larger cell size.

Defining a grid size $R_g = \Delta_x / \lambda_{De}$ the $\boldsymbol{k}_x$ and $\boldsymbol{k}_y$ integrals were evaluated numerically at velocities up to $3v_{Te}$. We would expect the grid aliasing effects to be most pronounced at larger grid sizes, so a comparison of the $R_g = 1$ case calculated with the first aliasing modes included $(p_x, p_y) \in [-1,1]$ was compared to the case with only the zeroth mode included $p_x = p_y = 0$. For both the spectral Maxwell solver and the FDTD Maxwell solver, under the grid scalings considered, the aliasing modes have a negligible effect on the drag and diffusion coefficients.

## Bibliography

[1] Z. Donko, A. Derzsi, M. Vass, B. Horvath, S. Wilczek, B. Hartmann and P. Hartmann, "eduPIC: an introductory particle based code for radio-frequency plasma simulation," *Plasma Sources Sci. Technol.,* vol. 30, p. 095017, 2021.

[2] S. Rauf, D. Sydorenko, S. Jubin, W. Villafana, S. Ethier, A. Khrabrov and I. Kaganovich, "Particle-in-cell modeling of electron beam generated plasma," *Plasma Sources Science and Technology,* vol. 32, no. 5, p. 05509, 2023.

[3] M. M. Turner, A. Derzsi, Z. Donko, D. Eremin, S. J. Kelly, T. Lafleur and T. Mussenbrock, "Simulation benchmarks for low-pressure plasmas: Capcitive discharges," *Physics of Plasmas,* vol. 20, p. 013507, 2013.

[4] M. Vass, P. Palla and P. Hartman, "Revisiting the numerical stability/accuracy conditions of explicit PIC/MCC simulations of low-temperature gas discharges," *Plasma Sources Sci. Technol.,* vol. 31, p. 064001, 2022.

[5] D.-Q. Wen, J. Krek, J. T. Gudmundsson, E. Kawamura, M. A. Lieberman and J. P. Verboncoeur, "Particle-in-cell simulations with fluid metastable atoms in capacitive argon discharges: Electron elastic scattering and plasma density profile transition," *IEEE Transactions on Plasma Science,* vol. 50, no. 9, pp. 2548-2557, 2022.

[6] O. C. Eldridge and M. Feix, "Numerical experiments with a plasma model," *The Physics of Fluids,* vol. 6, p. 398, 1963.

[7] J. -B. Fouvry, B. Bar-Or and P. -H. Chavanis, "Kinetic theory of one-dimensional homogeneous long-range interacting systems sourced by 1/N2 effects," *Physical Review E,* vol. 100, p. 052142, 2019.

[8] R. W. Hockney, "Measurements of collision and heating times in a two-dimensional thermal computer plasma," *Journal of Computational Physics,* vol. 8, pp. 19-44, 1971.

[9] D. Montgomery and C. W. Nielson, "Thermal relaxation in one- and two-dimensional plasma models," *The Physics of Fluids,* vol. 13, p. 1405, 1970.

[10] D. Sydorenko, A. Khrabrov, W. Villafana, S. Ethier and S. Janhunen, "EDIPIC-2D online repository," 2022. [Online]. Available: https://github.com/PrincetonUniversity/EDIPIC-2D.

[11] S. P. Gary, Y. Zhao, R. S. Hughes, J. Wang and T. N. Parashar, "Species entropies in the kinetic range of collisionless plasma turbulence: Particle-in-cell simulations," *The Astrophysical Journal,* vol. 859, p. 110, 2018.

[12] C. Birdsall and A. Langdon, Plasma Physics via Computer Simulation, Taylor and Francis, 2004.

[13] M. Touati, R. Codur, F. Tsung, V. K. Decyk, W. B. Mori and L. O. Silva, "Kinetic theory of particle-in-cell simulation plasma and the ensemble averaging technique," *Plasma Phys. Control. Fusion,* vol. 64, p. 115014, 2022.

[14] M. M. Turner, "Kinetic properties of particle-in-cell simulations compromised by Monte Carlo collisions," *Physics of Plasmas,* vol. 13, p. 033506, 2006.

[15] P. Y. Lai, T. Y. Lin, Y. R. Lin-Liu and S. H. Chen, "Numerical thermalization in particle-in-cell simulations with Monte-Carlo collisions," *Physics of Plasmas,* vol. 21, p. 122111, 2014.

[16] P. Y. Lai, L. Chen, Y. R. Lin-Liu and S. H. Chen, "Study of discrete-particle effects in a one-dimensional plasma simulation with the Krook type collision model," *Physics of Plasmas,* vol. 22, p. 092127, 2015.

[17] H. Okuda and C. K. Birdsall, "Collisions in a plasma of finite-size particles," *The Physics of Fluids,* vol. 13, p. 2123, 1970.

[18] B. Abraham-Shrauner, "Test particle in a two-dimensional plasma," *Physica,* vol. 43, pp. 95-104, 1969.

[19] M. A. Reynolds, B. D. Fried and G. J. Morales, "Velocity-space drag and diffusion in a model, two-dimensional plasma," *Phys. Plasmas,* vol. 4, no. 5, pp. 1286-1296, 1997.

[20] J. M. Dawson, "Thermal relaxation in a one-species, one-dimensional plasma," *The Physics of Fluids,* vol. 7, p. 419, 1964.

[21] J. Hsu, G. Joyce and D. Montgomery, "Thermal relaxation of a two-dimensional plasma in a d.c. magnetic field. Part 2. Numerical simulation," *J. Plasma Physics,* vol. 12, pp. 27-31, 1974.

[22] I. Jechart, T. Katsouleas and J. Dawson, "Anomalous thermal relaxation of a two-dimensional magnetized plasma," *The Physics of Fluids,* vol. 30, no. 1, p. 65, 1987.

[23] J. Virtamo and H. Tuomisto, "Verification of a simple collision operator for one-dimensional plasma by simulation experiments," *The Physics of Fluids,* vol. 22, p. 172, 1979.

[24] J. Dawson, "One-dimensional plasma model," *The Physics of Fluids,* vol. 5, p. 445, 1962.

[25] J. -Y. Hsu, D. Montgomery and G. Joyce, "Thermal relaxation of a two-dimensional plasma in a d.c. magnetic field. Part 1. Theory," *J. Plasma Physics,* vol. 12, pp. 22-26, 1974.

[26] N. Gatsonis and A. Spirkin, "A three-dimensional electrostatic particle-in-cell methodology on unstructured Delaunay-Voronoi grids," *Journal of Computational Physics,* vol. 288, pp. 3742-3761, 2009.

[27] S. Averkin and N. Gatsonis, "A parallel electrostatic Particle-in-Cell method on unstructured tetrahedral grids for large-scale bounded collisionless plasma simulations," *Journal of Computational Physics,* vol. 363, pp. 178-199, 2018.

[28] A. T. Powis and I. D. Kaganovich, "Accuracy of the explicit energy-conserving particle-in-cell method for under-resolved simulations of capacitively coupled plasma discharges," *Physics of Plasmas (accepted),* 2023 .

[29] H. Sun, S. Banarjee, S. Sharma, A. T. Powis, A. V. Khrabrov, D. Sydorenko, J. Chen and I. D. Kaganovich, "Direct implicit and explicit energy-conserving particle-in-cell methods for modeling of capacitively-coupled plasma devices," *Physics of Plasmas,* vol. 30, p. 103509, 2023.

[30] S. G. Walton, D. R. Boris, S. C. Hernandez, E. H. Lock, T. B. Petrova, G. M. Petrov, A. V. Jagtiani, S. U. Engelmann, H. Miyazoe and E. A. Joseph, "Electron beam generated plasmas: Characteristics and etching of silicon nitride," *Microelectronic Engineering,* vol. 168, pp. 89-96, 2017.

[31] S. G. Walton, D. R. Boris, S. G. Rosenberg, H. Miyazoe, E. A. Joseph and S. U. Engelmann, "Etching with electron beam-generated plasmas: Selectivity versus ion energy in silicon-based films," *J. Vac. Sci. Technol. A,* vol. 39, p. 033002, 2021.

[32] K. Nanbu, "Theory of cumulative small-angle collisions in plasmas," *Phys. Rev. E,* vol. 55, pp. 4642-4652, 1997.

[33] T. Charoy, J.-P. Boeuf, A. Bourdon, J. Carlsson, P. Chabert, B. Cuenot, D. Eremin, L. Garrigues, K. Hara, I. D. Kaganovich, A. T. Powis, A. Smolyakov, D. Sydorenko, A. Tavant, O. Vermorel and W. Villafana, "2D axial-azimuthal particle-in-cell benchmark for low-temperature partially magnetized plasmas," *Plasma Sources Science and Technology,* vol. 28, p. 105010, 2019.

[34] V. A. Godyak, R. B. Piejak and B. M. Alexandrovich, "Probe diagnostics of non-Maxwellian plasmas," *Journal of Applied Physics,* vol. 73, pp. 3657-3663, 1993.

[35] Y. He, Y.-M. Lim, J.-H. Lee, J.-H. Kim, M.-Y. Lee and C.-W. Chung, "Effect of parallel resonance on the electron energy distribution function in a 60 MHz capacitively coupled plasma," *Plasma Science and Technology,* vol. 25, p. 045401, 2023.

[36] I. D. Kaganovich and L. D. Tsendin, "The space-time-averaging procedure and modeling of the RF discharge, Part II: Motdel of collisional low-pressure RF discharge," *IEEE Transactions on Plasma Science,* vol. 20, no. 2, pp. 66-75, 1992.

[37] S. V. Berezhnoi, I. D. Kaganovich and L. D. Tsendin, "Generation of cold electrons. in a low-presssure RF capacitive discharge as an analogue of a thermal explosion," *Plasma Physics Reports,* vol. 24, pp. 556-563, 1998.

[38] K. Yee, "Numerical solution of initial boundary value problems involving Maxwell's equations in isotropic media," *IEEE Transactions on Antennas and Propagation,* vol. 14, no. 3, pp. 302-307, 1966.

[39] J. Villasenor and O. Buneman, "Rigorous charge conservation for local electromagnetic field solvers," *Computer Physics Communications,* vol. 69, pp. 306-316, 1992.

[40] PlasmaPy Community, *PlasmaPy, version 2023.5.1,* Zenodo, 2023.